\documentclass[12pt]{article}
\usepackage{graphicx}
\usepackage{epsfig}
\newcommand{\ccdot}{ \hspace{-0.8mm} \cdot \hspace{-0.8mm}}

\begin{document}

\title{{\bf Form factors in RQM approaches: \\ 
constraints from space-time translations, \\
extension to constituents with spin-1/2 \\ and unequal masses } } 
\author{ 
B.  Desplanques$^{1}$\thanks{{\it E-mail address:}  desplanq@lpsc.in2p3.fr},
Y.B. Dong$^{2,3}$\thanks{{\it E-mail address:}  dongyb@mail.ihep.ac.cn}
\\
$^{1}$LPSC, Universit\'e Joseph Fourier Grenoble 1, CNRS/IN2P3, INPG,\\
  F-38026 Grenoble Cedex, France \\
$^{2}$Institute of High Energy Physics, Chinese Academy of Science,\\ 
Beijing 100049, P. R. China\\
$^{3}$Theoretical Physics Center for Science Facilities (TPCSF), CAS,\\ 
Beijing 100049, P. R. China}

\sloppy

\maketitle

\begin{abstract}
\small{
\noindent
Constraints related to transformations of currents under space-time translations
have been considered for the relativistic quantum mechanics calculation 
of form factors of $J=0$ systems composed of scalar constituents 
with equal masses. Accounting for these constraints amounts 
to take into account many-body currents that restore the equality 
of the momentum transferred separately to the system and to the constituents, 
which holds in  field-theory approaches but is not generally fulfilled 
in relativistic quantum mechanics ones. When this was done, discrepancies between
results from different approaches could be found to vanish. The results are
extended here to systems composed of spin-1/2 constituents with unequal masses.
Moreover, as far as the equivalence of different approaches is concerned, some
intermediate step could be skipped and the presentation of these results
therefore slightly differs from the previous one. Due to the technical aspect of
present results, this work is not aimed to be published but it could be
useful for some applications like the form factors of the pion 
or kaon particles.}

\end{abstract} 
\newpage
\section{Introduction}
Constraints from properties related to space-time translations \cite{Lev:1993} 
are rarely mentioned in relativistic quantum mechanics (RQM) calculations 
of quantities like form factors. It was shown in previous works 
\cite{Desplanques:2004sp,Desplanques:2008fg,Desplanques:2008}
that accounting for these constraints could remove large discrepancies 
between different RQM calculations \cite{Dirac:1949cp,Keister:sb} 
as well as a dispersion-relation one 
\cite{Anisovich:1992hz,Melikhov:2001zv,Krutov:2001gu}. 
In some sense, accounting for these constraints 
is restoring the equality of the squared momentum transferred to
the constituents and to the whole system, as expected from a field-theory
approach, but violated in usual RQM approaches. It amounts to introduce
selected contributions from many-body currents at all orders 
in the interaction. These contributions compensate for interaction effects
implied by the choice of the hypersurface underlying some approach 
and allow one to restore the geometrical character of transformations 
under the Poincar\'e group. 
They can explain in particular two (related) paradoxes 
that both  point to missing some symmetry \cite{Desplanques:2003nk}. 
(1) In some cases, the charge radius could go to infinity 
when the mass of the system goes to zero. 
Generally, one expects the radius decreases when the 
attraction between constituents and therefore the binding energy, increases.
(2) The solution of a mass operator, which depends on internal variables, 
can be used independently of the way it has been derived. Arbitrary results for form
factors can then be obtained depending on the mass of the system.

Practically, to  account for the above
constraints, the coefficient of  the momentum transfer $q^{\mu}$ 
was multiplied  by a factor $\alpha$, which was determined  
so that to fulfill the equa\-lity of the square momentum transferred to the
constituents and to the  whole system, $(p_i-p_f)=q^2$. 
This factor was then expressed in terms of the spectator momentum $\vec{p}$. 
The arguments entering wave functions were also expressed in terms 
of this momentum and corrected to take into account the effect 
of the factor $\alpha$.  Form factors so obtained in one approach 
could thus be compared numerically with other ones. 
To show the algebraic identity of some corrected RQM approach with the
dispersion-relation approach or other RQM approaches, 
the momentum $\vec{p}$ was expressed in terms of the variables 
$s_i, \; s_f$ entering dispersion-relation expressions 
and a third quantity that could be integrated over.

The previous study was concerned with a system of two scalar constituents
with the same mass. Though this was not the primary motivation, 
this system offers the advantage that calculations can be compared 
to exact ones \cite{Amghar:2002jx}, based on solutions 
of the Bethe-Salpeter equation \cite{Wick:1954,Cutkosky:1954}. 
This system is somewhat academic however and there is therefore some need 
to extend the results obtained so far.
The extension can be done along two directions: systems with constituents of
different mass on the one hand, and with spin 1/2 on the other hand.
The  first extension is probably the most tedious one and
will  represent our main concern here.

While performing this new study, it appeared that the equation allowing one 
to get the factor $\alpha$ in terms of the momentum $\vec{p}$ 
could not be easily solved for some form as soon as the system 
was getting more sophisticated, with constituents of different masses 
for instance. However, it now appears that the whole dependence 
of the factor $\alpha$ on $\vec{p}$ goes exclusively through 
the variables $s_i, s_f$, without further dependence 
on the third quantity mentioned above. 
This allows one to simplify the demonstration showing that different approaches 
are algebraically equivalent but we lost in some cases (``point form") 
the possibility to check numerically this property directly 
from the original expressions of form factors corrected 
for the above mentioned constraints. 

As the motivation and the procedure for correcting form factors 
for taking into account constraints from space-time translations 
have been described at length in Ref. \cite{Desplanques:2008fg}, 
we will organize the present paper differently. Everything
concerned with a given approach is presented in a unique section. We are only
considering general cases with no special reference to Breit-frame results
previously obtained. As much as possible, the presentation in a given approach
parallels the one in another approach, so that to better emphasize the
similarities. This is especially true for the generalized instant- 
and front-form results.

The system under consideration consists of two constituents with unequal masses
in a $J=0$ state. For such a system there are two form factors, 
$F_0(Q^2)$ and $F_1(Q^2)$, respectively related to a scalar and a vector current
probe. Results are presented without regard to whether one is considering an
elastic or an inelastic transition. We assume that the vector current is a
conserved one. Getting results for spin-1/2 constituents can be done relatively
easily from the scalar-spin case. They amount to incorporate an extra factor in
the integrands entering the expression of form factors. This is done at the end
of each section dedicated to a given approach. Form factors can be defined in
different ways. In the present paper, we consider that only one of the
constituent (number 1, with mass $m_1$) interacts with the external field. 
They read:
\begin{eqnarray}
&&\sqrt{2E_f\;2E_i} \;\langle \;f\;|S_{op.}|\;i\; \rangle
= 4m_1 \;F_0(Q^2)\,,
\nonumber \\
&&\sqrt{2E_f\;2E_i} \;\langle \;f\;|J_{op.}^{\mu}|\;i\; \rangle
= (P_f+P_i)^{\mu} \;F_1(Q^2)\,,
\end{eqnarray}
where $F_0(Q^2)$ and $F_1(Q^2)$ are dimensionless quantities.  The above
expressions should be adapted for cases where the two constituents interact
with the external probe, possibly with different strengths. For constituents
with equal masses, their contributions are proportional to each other, allowing
one to factorize a common factor.

The plan of the paper is as follows. In the second section, 
we give expressions of the charge and scalar form factors 
in the dispersion-relation approach. The third section is devoted 
to the charge and scalar form factors in the front-form approach 
with the condition $q^+=0$. 
The fourth, fifth and sixth sections are concerned, respectively, 
with the instant \cite{Bakamjian:1953kh}, front and ``point-form" \cite{Bakamjian:1961,Sokolov:1978} approaches. 
They only involve the charge form factor. This one is considered 
in full generality, including arbitrary hyperplane orientations as well as
momentum configurations. Results for the spin-1/2 constituent case 
are given at the end of each section. Though we have in mind elastic form factors, 
results could be applied to some inelastic form factors 
in the space-like domain.

\begin{figure}[htb]
\begin{center}
\includegraphics[width=0.6\textwidth]{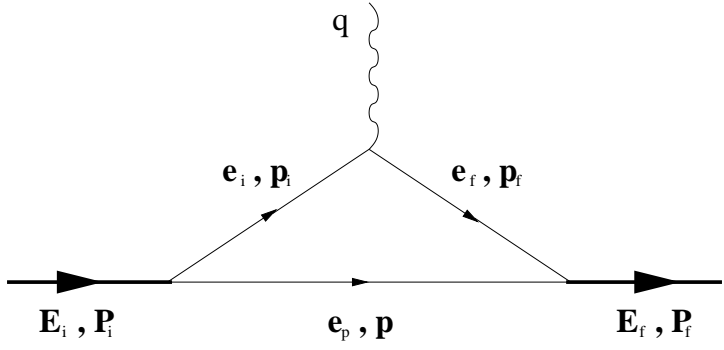}
\end{center}
\caption{Interaction  with an external field together with kinematical
definitions.\label{fig1}}
\end{figure}

The conventions used here are essentially the same as in a previous work
\cite{Desplanques:2008fg}. Kinema\-tical definitions relative to the interaction
with the external probe are reminded in Fig. \ref{fig1}. We  also use the
following conventions: 
$P_i-P_f=-q$, $\bar{P}=\frac{P_i+P_f}{2}$, $P_i=\bar{P}-\frac{q}{2}$, 
$P_f=\bar{P}+\frac{q}{2}$.  The notation $``\cdots"$ implies that 
what the dots account for incorporate effects of constraints
from space-time translations. The variables $s^0_{i,f}$ represent the quantities
$(p_{i,f}\!+\!p)^2$ uncorrected for the above effects (one has therefore 
$``s^0_{i,f}"=s_{i,f}$). The space- or time-like unit 4-vectors $\hat{q}^{\mu}$,
$\hat{\tilde{q}}^{\mu}$, $\hat{\lambda}^{\mu}$ and  $\hat{v}^{\mu}$ are defined
in the sections where they are used with the exception of $\hat{q}^{\mu}$, which
appears in different sections and is defined 
as $\hat{q}^{\mu}=q^{\mu}/\sqrt{-q^2}$.
%

\section{Incorporation of different masses (and spin 1/2) 
in the scalar-constituent results: dispersion-relation approach} 
We assume that $m_1$ represents the mass of the constituent interacting 
with the external field and $m_2$ the one for the spectator constituent.
We define the difference of the squared masses as $\Delta m^2=m_2^2-m_1^2$.

\noindent 
{\bf Function $I(s_i,Q^2,s_f)$}\\
We first give below details about the determination of the expression 
of the function $I(s_i,Q^2,s_f)$, which generalizes the equal-mass one 
given in Ref. \cite{Desplanques:2008fg}. 
We remind that $\tilde{P}_i^2=s_i$, $\tilde{P}_f^2=s_f$, 
$(\tilde{P}_i\!-\!\tilde{P}_f)^2=q^2=-Q^2$ and that the 4-vectors 
$\tilde{P}^{\mu}$ does not verify the usual on-mass-shell conditions 
($\tilde{P}^2 \neq M^2$). Noticing that 
$\int \frac{d\vec{p}_{i,f}}{2e_{i,f}}=\int d^4p_{i,f}\;\delta (p_{i,f}^2-m_1^2)$ 
(positive $p_0$), and $p_{i,f}^{\mu}=\tilde{P}_{i,f}^{\mu}\!-\!p^{\mu}$ 
(from the $\delta^4(\cdots)$ functions), 
the expression defining $I(s_i,Q^2,s_f)$   writes:
\begin{eqnarray}
&&I(s_i,Q^2,s_f)= \frac{1}{2\pi}\int \frac{d\vec{p}}{e_p}\;
 \delta(s_i\!+\!\Delta m^2\!-\!2 p \ccdot \tilde{P}_i)  \; 
 \delta(s_f\!+\!\Delta m^2\!-\!2 p \ccdot \tilde{P}_f ) \, .
\label{eq:dp1}
\end{eqnarray}
In order to make the integration over $\vec{p}$, we assume, 
without lost of generality, that the momenta, 
$\vec{\tilde{P}}_i, \;\vec{\tilde{P}}_f$, are in the $x,\,y$ plane.
As suggested by the above equation, we express the components of $\vec{p}$ 
in terms of $p \ccdot \tilde{P}_i $, $p \ccdot \tilde{P}_f $ and $p^z$.  
We thus obtain: 
\begin{eqnarray}
&&p^x =\frac {e_p(\tilde{P}_i^0\tilde{P}_f^y\!-\!\tilde{P}_f^0\tilde{P}_i^y)
-\Big( p \ccdot \tilde{P}_i \; \tilde{P}_f^y \!-\!p \ccdot \tilde{P}_f \; \tilde{P}_i^y \Big)
}{\tilde{P}_i^x\tilde{P}_f^y\!-\!\tilde{P}_i^y\tilde{P}_f^x } \, ,
\nonumber \\&&
p^y =\frac {e_p(\tilde{P}_i^0\tilde{P}_f^x\!-\!\tilde{P}_f^0\tilde{P}_i^x)
-\Big( p \ccdot \tilde{P}_i \; \tilde{P}_f^x \!-\!p \ccdot \tilde{P}_f \; \tilde{P}_i^x \Big)
}{\tilde{P}_i^y\tilde{P}_f^x\!-\!\tilde{P}_i^x\tilde{P}_f^y } \, , 
\label{eq:dp2}
\end{eqnarray}
where $e_p$, which is solution of a second-order equation, is given by: 
\begin{eqnarray}
e_p\!&\!=\!&\!\frac{1}{Q^2D}
\Bigg ((2\bar{s}\!+\!Q^2) (\tilde{P}_{i0}s_f\!+\! \tilde{P}_{f0}s_i) 
-2s_is_f(\tilde{P}_{i0}\!+\! \tilde{P}_{f0})
\nonumber \\ 
&&\hspace*{1.2cm} 
\pm 2Q \sqrt{ s_is_f\,c_{\Delta m^2} \!-\! (m_2^2\!+\!p^{z2})D}
\nonumber \\ 
&&\hspace*{4cm} \times \;\sqrt {\tilde{P}_{i0}\tilde{P}_{f0}\,(2\bar{s}\!+\!Q^2) 
\!-\! (\tilde{P}_{i0}^2s_f\!+\! {\tilde{P}_{f0}}^2s_i ) 
\!-\!\frac {Q^2D}{4}} \;\Bigg),\hspace*{0.2cm} 
\label{eq:dp3}
\end{eqnarray}
where $\bar{s}=\frac{s_i+s_f}{2}$. 
The quantity $D$ is the same as the one defined elsewhere 
\cite{Desplanques:2008fg}:
\begin{eqnarray}
D=
4\frac{(\tilde{P}_i \ccdot \tilde{P}_f)^2\!-\!\tilde{P}_i^2 \tilde{P}_f^2}{Q^2}
= 4\bar{s}+Q^2+\frac{(s_i\! -\! s_f)^2}{Q^2}\,,
\label{eq:dp4}
\end{eqnarray}
while the quantity $c_{\Delta m^2}$, introduced to simplify the writing of the
equation and future ones, is defined as:
\begin{eqnarray}
c_{\Delta m^2}=\Big(1\! +\! \frac{\Delta m^2}{s_i}\Big) \Big(1\! +\! \frac{\Delta m^2}{s_f}\Big)
+\frac{\Delta m^2(s_i\! -\! s_f)^2}{Q^2s_is_f}\, .
\label{eq:dp5}
\end{eqnarray}
The integration volume transforms as follows:
\begin{eqnarray}
\frac{d\vec{p}}{e_p}=\sum 
\frac{2\,d(p \ccdot \tilde{P}_i )\,d(p \ccdot \tilde{P}_f )\,dp^z \;
\theta(\cdots)}{   
   Q \sqrt { s_is_f\,c_{\Delta m^2}\!-\! (m_2^2\!+\!p^{z2})D}}\,,
   \label{eq:dp6}
\end{eqnarray}
where all the dependence on the components of the 4-vectors
$\tilde{P}^{\mu}_{i,f}$ is found to be absorbed 
into the quantities $s_i,\,s_f,\,Q^2$. The $\sum$ symbol accounts for the existence of the two solutions of $e_p$ given
in Eq. (\ref{eq:dp3}). The function $\theta(\cdots)$ 
is defined as: 
\begin{eqnarray}
\theta(\cdots)=\theta\Big(\frac{s_is_f}{D}\,c_{\Delta m^2}
-m_2^2\Big)\,. \label{eq:dp66}
\end{eqnarray}
It provides a minimal condition so that  the quantity under the square-root 
symbol at  the denominator in Eq. (\ref{eq:dp6}) be positive 
for some finite range of $p_z$
\footnote{We chose to incorporate the $\theta(\cdots)$ function at this stage
but one can imagine to incorporate it at a later stage, as lower and upper
limits in the integrals for instance.}. 

After inserting the last result into Eq. (\ref{eq:dp1}) and integrating 
over the variables $p \cdot \tilde{P}_i $,  $p \cdot \tilde{P}_f $, 
using the $\delta(\cdots)$ functions,
taking also into account that there are two solutions for $e_p$, 
one gets the desired result:
\begin{eqnarray}
I(\cdots)&\!=\!&\frac{1}{8\pi}\int dp^z  \sum
\frac{2\;\theta(\cdots)}{Q\sqrt {s_is_f\,c_{\Delta m^2}\!-\! (m_2^2\!+\!p^{z2})D}}
=\frac{\theta(\cdots)}{2Q\sqrt{D}}\,.\label{eq:dp7}
\end{eqnarray}
It is interesting to notice how the extra factor $\pi$ is obtained 
at the r.h.s.. 
In a system with azimuthal symmetry, a factor  proportional to $\pi$ simply 
arises from the integration over the azimuthal angle  $\phi$.
In the present case, where this symmetry is not assumed, it comes from an
integral of the type $\int^{|a|}_{-|a|}\frac{dz}{(a^2-z^2)^{1/2}}=\pi$.

\noindent 
{\bf Form factors (scalar constituents)}\\
For calculating scalar and charge form factors, we assumed that the interactions
of the constituents with the external probe are given by: 
\begin{eqnarray}
&&\tilde{S} \propto 2\,m_1\, ,
 \nonumber \\
&&\tilde{I}^{\mu} \propto (p_i+p_f)^{\mu}\,,
\label{eq:dp8}
\end{eqnarray}
where the proportionality factor contains the coupling constant and some
convention-dependent factor related to the definition. We assumed that the
interactions for the scalar and charge probes become the same in the
non-relativistic limit.
While the calculation of the scalar form factor is relatively easy, 
some care is required for the charge form factor so that to account for current
conservation. In this case,  one has to project the current 
$\tilde{I}^{\mu}$   on the 4-vector:
\begin{eqnarray}
A^{\mu}=p_i^{\mu}+p_f^{\mu}+2p^{\mu}- 
   \frac{(p_i\!+\!p_f\!+\!2p)\!\cdot\!(p_i\!-\!p_f)}{(p_i\!-\!p_f)^2}\, 
      (p_i^{\mu}-p_f^{\mu})\, ,  \label{eq:dp9}
\end{eqnarray}
and divide the result by the square, $A^2=A\!\cdot\!A$ 
where the 4-vector $A^{\mu}$
verifies the property of a conserved current, 
$q \ccdot A=(p_i-p_f)  \ccdot A=0$. 
The projection of the current on  $A^{\mu}$, defined as $\tilde{I}_a$, 
which enters the integrand for the charge form factor, is thus given by:
\begin{eqnarray}
&&\tilde{I}_a= \tilde{I}  \!\cdot\!  A \propto 
\Big(s_i \!+\!s_f \!-\!2 \Delta m^2 \!-\!q^2\Big)\, ,
\label{eq:dp10}
\end{eqnarray}
while the quantity  $A^2$ is related to the quantity D given 
in Eq. (\ref{eq:dp4}) by the relation:
\begin{eqnarray}
A^2=D=4\bar{s}+Q^2+\frac{(s_i-s_f)^2}{Q^2}\, .
\label{eq:dp11}
\end{eqnarray}
Dividing the integrand by this factor explains the appearance 
of an extra factor $D$ at the denominator for the charge form factor 
(in comparison with the scalar form factor).

One thus gets the following expressions for the charge and scalar 
form factors:
\begin{eqnarray} 
&&F_0(Q^2) = \frac{1}{N} \int d\bar{s} \;  
d(s_i\!-\!s_f)\; \phi(s_i) \; \phi(s_f) \; 
\frac{  \;\theta(\cdots) }{ 2\, 
\Big((s_i\!-\!s_f)^2\!+\!4Q^2\;\bar{s}\!+\!Q^4 \Big)^{1/2}\;} 
 \nonumber \\
&& \hspace*{1.3cm} =\frac{1}{N} \int d\bar{s} \;  d(\frac{s_i\!-\!s_f}{Q})\;
 \phi(s_i) \; \phi(s_f) \;\frac{ \theta(\cdots) }{
2\sqrt{D}\, }\, ,
 \nonumber \\
&&F_1(Q^2)=\frac{1}{N} \int d\bar{s} \;  
d(s_i\!-\!s_f)\; \phi(s_i) \; \phi(s_f) \; 
\frac{ \;Q^2\; [2\bar{s} \!-\!2 \Delta m^2 \!-\!q^2]\;\theta(\cdots) }{
\Big((s_i\!-\!s_f)^2\!+\!4Q^2\;\bar{s}\!+\!Q^4 \Big)^{3/2} }
 \nonumber \\
&& \hspace*{1.3cm} =\frac{1}{N} \int d\bar{s} \;  d(\frac{s_i\!-\!s_f}{Q}) \; 
\phi(s_i) \; \phi(s_f)\; \frac{ [2\bar{s} \!-\!2 \Delta m^2 \!+\!Q^2]
\;\theta(\cdots) }{D \sqrt{D}}\, . \label{eq:dp12}
\end{eqnarray}

\noindent 
{\bf Form factors (spin-1/2 constituents)}\\
When incorporating the spin-1/2 of the constituents, the quantities 
$\tilde{S}$ and $\tilde{I}^{\mu}$ have to be changed. 
One has to account for the spin wave function with its appropriate normalization,
$\bar{u}(p_{i,f}) \,\gamma_5 \, {v}(p)/\sqrt{s_{i,f}-(m_1\!-\!m_2)^2}$, 
and the current expressions, $\bar{u}(p_{i})\, u(p_{f}) $ or 
$\bar{u}(p_{i}) \, \gamma^{\mu} \, u(p_{f}) $. To obtain the factors to be
inserted in the integrands for form factors, the trace on Dirac matrices 
has to be performed. One gets: 
\begin{eqnarray}
&&S\propto \frac{2m_1\Big(\bar{s}\!-\!(m_1\!-\!m_2)^2\Big)-m_2\,q^2     }{  
\sqrt{s_i-(m_1\!-\!m_2)^2}\;\sqrt{s_f-(m_1\!-\!m_2)^2} }\, ,
\nonumber \\
&&I^{\mu} \propto \frac{p_i^{\mu}\Big(s_f\!-\!(m_1\!-\!m_2)^2\Big)
+ p_f^{\mu}\Big(s_i\!-\!(m_1\!-\!m_2)^2\Big)+p^{\mu}q^2    }{  
\sqrt{s_i-(m_1\!-\!m_2)^2}\;\sqrt{s_f-(m_1\!-\!m_2)^2}}\, ,
\nonumber \\
&&I_a= I  \!\cdot\! A \propto 
\frac{2s_i\,s_f-\Delta m^2(2\bar{s}  \!-\!q^2)
-(m_1\!-\!m_2)^2(2\bar{s} \!-\!2 \Delta m^2 \!-\!q^2)
}{ \sqrt{s_i-(m_1\!-\!m_2)^2}\;\sqrt{s_f-(m_1\!-\!m_2)^2}} \, ,\label{eq:dp13}
\end{eqnarray}
where the proportionality factor is the same 
as in Eqs. (\ref{eq:dp8}, \ref{eq:dp10} ). Expressions for form factors can be
simply obtained by multiplying the integrands for the scalar-constituent case
in Eqs. (\ref{eq:dp11}) by the factors:
\begin{eqnarray}
&& \frac{S}{\tilde{S}}=\frac{ 2m_1\Big(\bar{s}\!-\!(m_1\!-\!m_2)^2\Big)-m_2\,q^2   }{
2m_1\sqrt{s_i-(m_1\!-\!m_2)^2}\;\sqrt{s_f-(m_1\!-\!m_2)^2}}
\, ,
\nonumber \\
&& \frac{I_a}{\tilde{I}_a}=
 \frac{2s_i\,s_f-\Delta m^2(2\bar{s}  \!-\!q^2)
-(m_1\!-\!m_2)^2( 2\bar{s} \!-\!2 \Delta m^2 \!-\!q^2)
}{\Big(2\bar{s} \!-\!2 \Delta m^2 \!-\!q^2\Big) 
\sqrt{s_i-(m_1\!-\!m_2)^2}\;\sqrt{s_f-(m_1\!-\!m_2)^2}} \, .\label{eq:dp14}
\end{eqnarray}
Introducing the above ratios in the expressions of form factors for scalar
constituents given by Eq. (\ref{eq:dp12}), one gets:
\begin{eqnarray} 
&&F_0(Q^2) = \frac{1}{N} \int d\bar{s} \;  d(\frac{s_i\!-\!s_f}{Q})\;
 \phi(s_i) \; \phi(s_f) 
\nonumber \\ 
&&\hspace*{2cm} \times \frac{\Big[2m_1
\Big(\bar{s}\!-\!(m_1\!-\!m_2)^2\Big)+m_2\,Q^2\Big] \;\theta(\cdots) }{
2\sqrt{D}\,(2m_1)\,\sqrt{s_i-(m_1\!-\!m_2)^2}\;\sqrt{s_f-(m_1\!-\!m_2)^2} } \, ,
\nonumber \\
&&F_1(Q^2) =\frac{1}{N} \int d\bar{s} \;  d(\frac{s_i\!-\!s_f}{Q}) \; 
 \phi(s_i) \; \phi(s_f)\;
 \nonumber \\
&& \hspace*{2cm}\times\frac{\Big[ 2s_i\,s_f\!-\!\Delta m^2(2\bar{s}  \!+\!Q^2)
\!-\!(m_1\!-\!m_2)^2(2\bar{s}\!-\!2 \Delta m^2 \!+\!Q^2) \Big]
\;\theta(\cdots) }{
D\sqrt{D}\sqrt{s_i-(m_1\!-\!m_2)^2}\;\sqrt{s_f-(m_1\!-\!m_2)^2}} \, .
\label{eq:dp15}
\end{eqnarray}
We notice that the above expression for the charge form factor agrees 
with the one given in Refs. \cite{Anisovich:1992hz,Melikhov:2001zv} 
but disagrees with the one given in Ref. \cite{Krutov:2001gu} 
for equal-mass constituents.
The discrepancy factor in the integrand, $(s_i+s_f+Q^2)/(2 \sqrt{s_i\;s_f})$, 
is the same as the factor found for scalar constituents \cite{Desplanques:2008fg}.
In this case, expressions of form factors, Eqs. (\ref{eq:dp12}), 
were checked by considering the simplest Feynman 
triangle diagram,  including unequal constituent masses or different masses for 
the initial and final states.

\noindent 
{\bf Normalization}\\
The integration over $s_i\!-\!s_f$ in the expression of $F_1(Q^2)$ can be performed
in the limit $Q^2=0$. The integration limit is given by the successive set 
of equations:
\begin{eqnarray} 
&&(s_i\!+\!\Delta m^2)(s_f\!+\!\Delta m^2)
\!+\! \frac{(s_i\!-\!s_f)^2}{Q^2}\Delta m^2\!-\! m_2^2\,D\geq 0 \, ,
 \nonumber \\
&&\bar{s}^2-2\bar{s}\,(m_2^2\!+\!m_1^2)-
(s_i\!-\!s_f)^2 \,(\frac{m_1^2}{Q^2}+\frac{1}{4})
+(m_2^2\!-\!m_1^2)^2-m_2^2\,Q^2\geq 0 \, ,
 \nonumber \\
&&|\frac{s_i\!-\!s_f}{Q}| \leq 
\sqrt{\frac{\bar{s}^2-2\bar{s}\,(m_2^2\!+\!m_1^2)+(m_2^2\!-\!m_1^2)^2}{m_1^2}}\, ,
\label{eq:dp16}
\end{eqnarray}
where only the most singular terms in the limit $Q\rightarrow 0$ 
have been retained at the last line. In this limit, the integration 
over $|\frac{s_i\!-\!s_f}{Q}|$ in the expression of  $F_1(Q^2)$ 
can be performed:
\begin{eqnarray} 
&&F_1(Q^2)_{Q\rightarrow  0}=\frac{1}{N} \int d\bar{s} \;  d(\frac{s_i\!-\!s_f}{Q}) 
\; \phi(s_i) \; \phi(s_f)\; 
\frac{ [2\bar{s} \!-\!2 \Delta m^2 \!-\!q^2]\;\theta(\cdots) }{
D^{3/2} }
\nonumber \\
&&\hspace*{1cm}=\frac{1}{N} \int d\bar{s} \; 2(\bar{s}\!+\!m_1^2\!-\!m_2^2)\; 
\phi^2(\bar{s}) \,\frac{2}{4\bar{s}} 
\frac{\sqrt{\bar{s}^2-2\bar{s}\,(m_2^2\!+\!m_1^2)+(m_2^2\!-\!m_1^2)^2}}{
\sqrt{\bar{s}^2-2\bar{s}\,(m_2^2\!+\!m_1^2)+(m_2^2\!-\!m_1^2)^2+4\bar{s}m_1^2}}
\nonumber \\
&&\hspace*{1cm}= \frac{1}{N} \int d\bar{s} \;\phi^2(\bar{s}) \,
\frac{\sqrt{\bar{s}^2-2\bar{s}\,(m_2^2\!+\!m_1^2)+(m_2^2\!-\!m_1^2)^2}
}{\bar{s}} \,. \label{eq:dp17}
\end{eqnarray}
It is noticed that the above expression is symmetrical in the exchange of the
constituent masses, $m_1$ and $m_2$, as expected.

Some relation to the expression of the norm in terms 
of the internal variable $k$ can be obtained as follows. 
Using the Bakamjian-Thomas transformation for unequal constituent masses
\cite{Bakamjian:1953kh},  
possibly generalized to any form \cite{Desplanques:2004sp} 
\footnote{Factors $e_k$ in Eq. (2) of this reference should be replaced 
by $e_{1k,2k}$ depending on the particle and the factor $2e_k$  
in the  next equations (3, 4, 5) should be replaced by $e_{1k}+e_{2k}$.}, 
one can express the $s$ variable entering the wave function $\phi(s)$ as:
\begin{eqnarray} 
s=(p_1+p_2)^2=(e_{1k}+e_{2k})^2\, ,
\end{eqnarray}
where $e_{1k}=\sqrt{m_1^2+k^2}$, $e_{2k}=\sqrt{m_2^2+k^2}$. The wave function 
$\tilde{\phi}(k^2)$ that is useful for our purpose is then given by:
\begin{eqnarray} 
\phi(s)=
\tilde{\phi}(\frac{s^2-2s\,(m_2^2\!+\!m_1^2)+(m_2^2\!-\!m_1^2)^2}{4s})
= \tilde{\phi}(k^2)\, .
\end{eqnarray}
Noticing that the above expression for $s$ implies relations such as:
\begin{eqnarray} 
&&2k=\frac{\sqrt{s^2-2s\,(m_2^2\!+\!m_1^2)+(m_2^2\!-\!m_1^2)^2}
}{\sqrt{s}}\, , \hspace*{0.5cm}
\nonumber \\
&&ds=\frac{2k\,(e_{1k}\!+\!e_{2k})^2}{e_{1k}\,e_{2k}}\, dk \,,
\end{eqnarray}
the expression of the norm given by Eq. (\ref{eq:dp17}) can be cast into 
the following one in terms of the $k$ variable:
\begin{eqnarray} 
&&F_1(0)=\frac{8}{N} \int dk\, k^2\, \tilde{\phi}^2(k^2) \;
\frac{e_{1k}\!+\!e_{2k}}{2\,e_{1k}\,e_{2k}}\,. \label{eq:dp18}
\end{eqnarray}
This last expression is a rather straigthforward generalization of the norm 
for equal-mass constituents.
\section{Incorporation of different masses (and spin 1/2) 
in the scalar-constituent results: front-form with $q^{+}=0$} 
We consider in this section a version of the front-form approach 
that can be obtained from a general one when $q \cdot \omega=q^+=0$. 
For this case, the properties from transformations of currents 
under space-time translations, which imply the equality of the square momentum 
transferred to the constituents and to the whole system, are trivially
fulfilled.  The factor $\alpha$ that was introduced elsewhere  for fulfilling
the above constraints \cite{Desplanques:2008fg} verifies the relation 
$\alpha=1$. No correction to the calculation of form factors is therefore 
needed in this case for accounting for the above properties.
It is then convenient to use 
the Bj\"orken variable $x=\frac{p\cdot\omega}{\bar{P}\cdot\omega}$ 
and components of momenta perpendicular to the front orientation, 
supposed to be in the direction of the $z$ axis.
Standard expressions of form factors to start with could be:
\begin{eqnarray} 
F_0(Q^2) &\! =\! & \frac{1}{\pi\,N} 
\int\! d^2R \int_0^1 \!\frac{dx}{2x(1\!-\!x)^2} \;  
\phi(s_i) \; \phi(s_f)  \;,
\nonumber \\ 
F_1(Q^2) & \!=\! & \frac{1}{\pi\,N} 
\int \! d^2R \int_0^1\! \frac{dx}{x(1\!-\!x)} \;  
\phi(s_i) \; \phi(s_f)   \;,
\label{eq:lf1} 
\end{eqnarray} 
where the arguments, $s_i$ and $s_f$, entering the wave functions may be written as:
\begin{eqnarray}
&&s_i=(p\!+\!p_i)^2=
\frac{m_1^2\!+\!p^2_{i\perp}}{1-x}+\frac{m_2^2\!+\!p^2_{\perp}}{x}-P^2_{i\perp}=
\frac{x\,m_1^2+(1\!-\!x)m_2^2
+(\vec{R}\!-\!x\,\vec{P}_{i\perp})^2}{x\,(1-x)}\, , 
\nonumber \\ 
&&s_f=(p\!+\!p_f)^2=
\frac{m_1^2\!+\!p^2_{f\perp}}{1-x}+\frac{m_2^2\!+\!p^2_{\perp}}{x}-P^2_{f\perp}=
\frac{x\,m_1^2+(1\!-\!x)m_2^2+(\vec{R}\!-\!x\vec{P}_{f\perp})^2}{x\,(1-x)} \, .
\nonumber \\ \label{eq:lf2} 
\end{eqnarray} 
In these last expressions,  $\vec{P}_{i\perp}$ and  $\vec{P}_{f\perp}$ 
are  two-dimensional vector representing the components 
of the initial or final  total momentum perpendicular to the front orientation 
(along the $z$ axis). The equality on the right is obtained by writing  
$\vec{p}_{\perp}=\vec{R}$ and $\vec{p}_{i\perp}=\vec{P}_{i\perp}-\vec{R}$ 
and  $\vec{p}_{f\perp}=\vec{P}_{f\perp}-\vec{R}$. 

\noindent
{\bf Expressions of $\vec{R}$ in terms of $\bar{s}$, $s_i\!-\!s_f$ and $x$.}\\
By making the change of variable 
$\vec{R}\!-\!x\,\vec{P}_{i\perp} \rightarrow \vec{R}$, one can get expressions
that only depend on $\vec{P}_{i\perp}-\vec{P}_{f\perp}=-\vec{Q}_{\perp}$,
where the two-dimensional vector $\vec{Q}_{\perp}$ represents the component 
of the  momentum transfer perpendicular to the same front orientation 
(notice that $Q^0+Q^z=0$, so that 
$Q^2=\vec{Q}^2_{\perp}+(Q^z)^2-(Q^0)^2=\vec{Q}_{\perp}^2$).
For our purpose, we make a ``symmetrical choice"
$\vec{R}\!-\!x\,\vec{P}_{i\perp} \rightarrow 
\vec{R}\!+\!x\,\frac{\vec{Q}_{\perp}}{2}$ and, consequently,
$\vec{R}\!-\!x\,\vec{P}_{f\perp} \rightarrow 
\vec{R}\!-\!x\,\frac{\vec{Q}_{\perp}}{2}$. 
Arguments entering the wave functions now read: 
\begin{eqnarray}
&& s_i= \frac{x\,m_1^2+(1\!-\!x)\,m_2^2
+\vec{R}^2\!+\!x\,\vec{R}\!\cdot\! \vec{Q}_{\perp} \!+\!\frac{x^2Q^2}{4}}{x\,(1-x)}\, , 
\nonumber \\ 
&& s_f= \frac{x\,m_1^2+(1\!-\!x)\,m_2^2
+\vec{R}^2\!-\!x\,\vec{R}\!\cdot\! \vec{Q}_{\perp} \!+\!\frac{x^2Q^2}{4}}{x\,(1-x)}\, ,
 \label{eq:lf3}
\end{eqnarray} 
from which we obtain:
\begin{eqnarray}
&& \bar{s}=\frac{x\,m_1^2+(1\!-\!x)\,m_2^2
     +\vec{R}^2\! +\!\frac{x^2Q^2}{4}}{x\,(1-x)}\, ,
\nonumber \\ 
&&s_i\!-\!s_f=2\,\frac{\vec{R}\!\cdot\! \vec{Q}_{\perp}}{1-x}\, .
 \label{eq:lf5}
\end{eqnarray} 
These equations can be inverted to express the components of $\vec{R}$ 
in terms of $\bar{s}$, $s_i\!-\!s_f$ and $x$. Assuming $\vec{Q}_{\perp}$ 
along the $x$ axis, one gets:
\begin{eqnarray}
&& R_x=(1\!-\!x)\,\frac{s_i\!-\!s_f}{2Q}\, ,
\nonumber \\ 
&& R_y=\pm \sqrt{x\,(1\!-\!x)\,\bar{s}-x\,m_1^2\!-\!(1\!-\!x)\,m_2^2-\frac{x^2Q^2}{4}
- (1\!-\!x)^2\frac{(s_i\!-\!s_f)^2}{4Q^2}} \, .
 \label{eq:lf6}
\end{eqnarray} 
\noindent
{\bf Jacobian for the variable transformation}\\
The Jacobian for the transformation of variables 
$\vec{R},\; x\rightarrow  s_i,\;s_f,\;x$ can be obtained from the relations:
\begin{eqnarray}
&&\frac{dR_x}{d(s_i\!-\!s_f)}=\frac{(1\!-\!x)}{2Q}\, ,
\nonumber \\ 
&&\frac{dR_y}{d\bar{s}}=\frac{\pm x\,(1\!-\!x) }{
2\sqrt{x\,(1\!-\!x)\,\bar{s}-x\,m_1^2\!-\!(1\!-\!x)\,m_2^2-\frac{x^2Q^2}{4}
- (1\!-\!x)^2\frac{(s_i\!-\!s_f)^2}{4Q^2}}}\, ,
 \label{eq:lf7}
\end{eqnarray} 
To go further, it is appropriate to rewrite the square-root factor appearing in
the above expression:
\begin{eqnarray}
&&2\sqrt{x\,(1\!-\!x)\,\bar{s}-x\,m_1^2\!-\!(1\!-\!x)\,m_2^2-\frac{x^2Q^2}{4}
- (1\!-\!x)^2\frac{(s_i\!-\!s_f)^2}{4Q^2}}
\nonumber \\ 
&&\hspace*{2cm}=\sqrt{D} \sqrt{ \Big  (\frac {s_i\,s_f }{ D }\,c_{\Delta m^2}-m_2^2\Big )f 
    \!-\!\Big (x\!-\! d \Big)^2 }\, ,
 \label{eq:lf8}
\end{eqnarray} 
where:
\begin{eqnarray}
d=\frac{2\bar{s}\!+\!2\Delta m^2\!+\!\frac{(s_i\!-\!s_f)^2}{Q^2}}{D}
=1\!-\!\frac{2\bar{s}\!-\!2\Delta m^2\!+\!Q^2}{D}
\,,\hspace*{1cm} f=\frac{4}{D}\,.
 \label{eq:lf77}
\end{eqnarray} 
One gets for the integration volume:
\begin{eqnarray}
d^2R\;dx=
\frac{2 x\,(1\!-\!x)^2 \,d\frac{(s_i\!-\!s_f)}{Q}\,d\bar{s}\;dx\; \theta(\cdots)}{
2\sqrt{D} \sqrt{ \Big  (\frac {s_i\,s_f}{ D }\,c_{\Delta m^2}-m_2^2\Big )f 
    \!-\!\Big (x\!-\! d \Big)^2 }}\, ,
 \label{eq:lf78}
\end{eqnarray} 
where a factor 2 has been introduced at the numerator to  account 
for the two solutions with opposite signs in Eq. (\ref{eq:lf7}).  
The above quantity makes sense provided that the quantity multiplying $f$ 
under the square-root symbol is positive. To account for this fact, 
we also introduce at the numerator a $\theta(\cdots)$ function, 
which is the same  as the one defined  in Eq. (\ref{eq:dp66}). 

\noindent
{\bf Form factors (scalar constituents)}\\
The equality with the dispersion-relation results can be shown by expressing 
the integrals in Eqs. (\ref{eq:lf1}) in terms of the variables
entering  a dispersion-type approach, $s_i$ and $s_f$, together with $x$, 
and integrating over $x$. Beginning with the simplest case of the scalar form
factor, one gets: 
\begin{eqnarray} 
F_0(Q^2) & \!=\! & \frac{1}{\pi N} 
\int \!  d\bar{s} \;  d(\frac{s_i\!-\!s_f}{Q}) \;
\phi(s_i) \; \phi(s_f)\; \theta(\cdots)
\nonumber \\
&&\hspace*{1cm}\times  \int_{x^-}^{x^+}\! \frac{dx \;}{ 2 \sqrt{D}\,
\sqrt{ \Big  (\frac {s_i\,s_f }{ D }\,c_{\Delta m^2}-m_2^2\Big )f \;  
   \!-\!\Big (x\!-\! d \Big)^2 } }
\nonumber \\
& \!=\! & \frac{1}{ N} \int \!  d\bar{s} \;  d(\frac{s_i\!-\!s_f}{Q})\;
 \phi(s_i) \; \phi(s_f) \; \frac{\theta(\cdots)}{2\sqrt{D}} \;,
 \label{eq:lf9}
\end{eqnarray} 
where ${x^\pm}$ represent the upper and lower values of the $x$ variable 
that cancel the quantity under the square-root symbol 
at the denominator in the second line. The derivation of the expression 
for the charge form factor is slightly more complicated:
\begin{eqnarray} 
F_1(Q^2) & \!=\! & \frac{1}{\pi N} 
\int \!  d\bar{s} \;  d(\frac{s_i\!-\!s_f}{Q}) \;
\phi(s_i) \; \phi(s_f)\;\theta(\cdots)
\nonumber \\
&&\hspace*{1cm}\times  \int_{x^-}^{x^+} \! \frac{dx \;(1\!-\!x)}{  
 \sqrt{D} \,
\sqrt{ \Big  (\frac {s_i\,s_f }{ D }\,c_{\Delta m^2}-m_2^2\Big )f \;  
   \!-\!\Big (x\!-\! d \Big)^2 } } \;  
\nonumber \\
 & \!=\! & \frac{1}{ N} \int \!  d\bar{s} \;  d(\frac{s_i\!-\!s_f}{Q}) \;
 \phi(s_i) \; \phi(s_f)\;
 \frac{ [ 2\bar{s}\!-\!2\Delta m^2\!+\!Q^2]\;\theta(\cdots)}{D\sqrt{D}}\;.
 \label{eq:lf10}
\end{eqnarray} 
The result at the last line has been obtained by writing:
\begin{eqnarray} 
1\!-\!x=\frac{ 2\bar{s}\!-\!2\Delta m^2\!+\!Q^2}{D}
- \Big(x\!-\!d\Big)\, ,
 \label{eq:lf11}
\end{eqnarray} 
and skipping the second term which gives 0 upon integration over $x$.

\noindent
{\bf Form factors (spin-1/2 constituents)}\\
The ratios correcting for the spin-1/2 nature of the constituents are given by:
\begin{eqnarray}
&& \frac{S}{\tilde{S}}=\frac{ 2m_1\Big(\bar{s}\!-\!(m_1\!-\!m_2)^2\Big)\!-\!m_2\,q^2   }{
2m_1\sqrt{s_i\!-\!(m_1\!-\!m_2)^2}\;\sqrt{s_f\!-\!(m_1\!-\!m_2)^2}}\, ,
\nonumber \\
&& \frac{I^0_{\omega}}{\tilde{I}^0_{\omega}}=
 \frac{2(1\!-\!x)\Big(\bar{s}\!-\!(m_1\!-\!m_2)^2\Big)+xq^2
}{2(1\!-\!x) \sqrt{s_i\!-\!(m_1\!-\!m_2)^2}\;\sqrt{s_f\!-\!(m_1\!-\!m_2)^2}}\,,
 \label{eq:lf12}
\end{eqnarray}
where $I^0_{\omega}, \; \tilde{I}^0_{\omega}$ in the second equation 
are defined similarly to Eq. (\ref{eq:dp10}), with $A^{\mu}$ 
replaced by $\omega^{\mu}$. The superscript $0$  reminds that $q^+=0$.
Introducing the above ratios in Eqs. (\ref{eq:lf10}), one gets the expressions
of form factors for  spin-1/2 constituents:
\begin{eqnarray} 
F_0(Q^2) & \!=\! & \frac{1}{\pi N} 
\int \!  d\bar{s} \;  d(\frac{s_i\!-\!s_f}{Q}) \;
\phi(s_i) \; \phi(s_f)\;\frac{ \Big(2m_1\Big(\bar{s}\!-\!(m_1\!-\!m_2)^2\Big)\!-\!m_2\,q^2  
 \Big)\;\theta(\cdots)}{2m_1 \sqrt{s_i\!-\!(m_1\!-\!m_2)^2}\;\sqrt{s_f\!-\!(m_1\!-\!m_2)^2} }
\nonumber \\
&& \times  \int_{x^-}^{x^+}\! \frac{dx }{ 2 \sqrt{D}\,
\sqrt{ \Big  (\frac {s_i\,s_f }{ D }\,c_{\Delta m^2}-m_2^2\Big )f \;  
   \!-\!\Big (x\!-\! d \Big)^2 } }
\nonumber \\
& \!=\! & \frac{1}{ N} \int \!  d\bar{s} \;  d(\frac{s_i\!-\!s_f}{Q})\; 
\phi(s_i) \; \phi(s_f)\;
\frac{\Big[ 2m_1\Big(\bar{s}\!-\!(m_1\!-\!m_2)^2\Big)\!+\!m_2\,Q^2\Big]\;
\theta(\cdots)      }{   2\sqrt{D} (2m_1) 
\sqrt{s_i\!-\!(m_1\!-\!m_2)^2}\;\sqrt{s_f\!-\!(m_1\!-\!m_2)^2}  }  \;,
\nonumber \\
F_1(Q^2) & \!=\! & \frac{1}{\pi N} 
\int \!  d\bar{s} \;  d(\frac{s_i\!-\!s_f}{Q}) \;
\phi(s_i) \; \phi(s_f)\;
\frac{\theta(\cdots)}{\sqrt{s_i\!-\!(m_1\!-\!m_2)^2}\;\sqrt{s_f\!-\!(m_1\!-\!m_2)^2}}
\nonumber \\
&& \times  \int_{x^-}^{x^+} \! 
\frac{dx \;  \Big  [ 2(1\!-\!x)\Big(\bar{s}\!-\!(m_1\!-\!m_2)^2\Big)\!+\!xq^2\Big ]    
  }{   2\sqrt{D} \,
\sqrt{ \Big  (\frac {s_i\,s_f }{ D }\,c_{\Delta m^2}-m_2^2\Big )f \;  
   \!-\!\Big (x\!-\! d \Big)^2 } } \;  
\nonumber \\
 & \!=\! & \frac{1}{ N} \int \!  d\bar{s} \;  d(\frac{s_i\!-\!s_f}{Q})\;
 \phi(s_i) \; \phi(s_f)
\nonumber \\
&& \times
 \frac{ \Big [2s_is_f\!-\!\Delta m^2(2\bar{s}\!+\!Q^2) 
\!-\!(m_1\!-\!m_2)^2 (2\bar{s}\!-\!2\Delta m^2\!+\!Q^2)\Big]\;\theta(\cdots)}{
D\sqrt{D}\sqrt{s_i\!-\!(m_1\!-\!m_2)^2}\;\sqrt{s_f\!-\!(m_1\!-\!m_2)^2} }
 \;.
 \label{eq:lf13}
\end{eqnarray} 
The result at the last line has been obtained by writing:
\begin{eqnarray} 
&& 2(1\!-\!x)(\bar{s}\!-\!(m_1\!-\!m_2)^2)\!+\!xq^2 
\nonumber \\
&&\hspace*{1.2cm}=
\Big[ 2(1\!-\!d)(\bar{s}\!-\!(m_1\!-\!m_2)^2)\!+\!dq^2\Big]
      - (x\!-\!d)\Big[2\bar{s}\!-\!2(m_1\!-\!m_2)^2\!-\!q^2\Big]
 \nonumber \\
&&\hspace*{1.2cm}=2\frac{\Big [2s_is_f\!-\!\Delta m^2(2\bar{s}\!+\!Q^2) 
\!-\!(m_1\!-\!m_2)^2 (2\bar{s}\!-\!2\Delta m^2\!+\!Q^2)\Big]}{D} 
 \nonumber \\
&&\hspace*{7.2cm} - 2(x\!-\!d)\Big[\bar{s}\!-\!(m_1\!-\!m_2)^2\!+\!
\frac{Q^2}{2}\Big]\, ,
 \label{eq:lf14}
\end{eqnarray} 
and skipping the last term which gives 0 upon integration over $x$.
\section{Incorporation of different masses (and spin 1/2) 
in the scalar-constituent results: generalized hyperplane} 
The relation to start with in the generalized hyperplane case  with orientation
$\lambda^{\mu}$ (finite
$\lambda^2$) is given by Eq. (16) of Ref. \cite{Desplanques:2008fg}:
\begin{eqnarray} 
p_{i,f}^{\mu}&\!=\!&P_{i,f}^{\mu}-p^{\mu}
+\frac{\lambda^{\mu}}{\lambda^2}
\Bigg(
\sqrt{ (\lambda\! \cdot\! P_{i,f})^2\!+\! (s^0_{i,f}\!-\!P_{i,f}^2)\,\lambda^2}
-\lambda\! \cdot\! P_{i,f}\Bigg)\, .
 \label{eq:if1}
\end{eqnarray}
{\bf Derivation of $\alpha$:}\\
One has the following set of equations:
\begin{eqnarray} 
p_i^{\mu}-p_f^{\mu}&\!=\!&P_i^{\mu}\!-\!P_f^{\mu}
\nonumber \\
&&+\frac{\lambda^{\mu}}{\lambda^2}
\Bigg(\sqrt{ (\lambda\! \cdot\! P_i)^2\!+\! (s^0_i\!-\!P_i^2)\,\lambda^2}
-\sqrt{ (\lambda\! \cdot\! P_f)^2\!+\! (s^0_f\!-\!P_f^2)\,\lambda^2}
- \lambda\! \cdot\! (P_i\!-\!P_f) \Bigg)\, ,
\nonumber \\
(p_i\!-\!p_f)^2&\!=\!& (P_i\!-\!P_f)^2
-\frac{(\lambda\! \cdot\! (P_i\!-\!P_f))^2 }{\lambda^2} 
\nonumber \\
&&+ \frac{\Bigg(\sqrt{ (\lambda\! \cdot\! P_i)^2\!+\! (s^0_i\!-\!P_i^2)\,\lambda^2}
-\sqrt{ (\lambda\! \cdot\! P_f)^2\!+\! (s^0_f\!-\!P_f^2)\,\lambda^2}\Bigg)^2}{
\lambda^2}
\nonumber \\
&\!=\!& (\tilde{P}_i\!-\!\tilde{P}_f)^2
+\Bigg(\sqrt{  s^0_i\!-\!\tilde{P}_i^2}
-\sqrt{ s^0_f\!-\!\tilde{P}_f^2}\Bigg)^2\, ,
 \label{eq:if2}
\end{eqnarray}
where we introduced at the last line the notation 
$\tilde{P}_{i,f}^{\mu}=P_{i,f}^{\mu}-\lambda^{\mu}\;(\lambda \cdot P_{i,f})/\lambda^2$.  
In the following, and in order of  simplifying the
writing of some quantities, we also use the notations: 
$\tilde{q}^{\mu}=q^{\mu}-\lambda^{\mu}\;(\lambda \cdot q)/\lambda^2$ 
and $\hat{\tilde{q}}^{\mu}= \tilde{q}^{\mu}/\sqrt{-\tilde{q}^2}$, 
$Q=\sqrt{-q^2}$, $\hat{\lambda}^{\mu}= \lambda^{\mu}/\sqrt{\lambda^2}$, 
as well as relations like  
$\tilde{P}_{i,f} \cdot \tilde{q}= P_{i,f} \cdot \tilde{q}$. 

Accounting for corrections related to space-time translation properties, 
one gets the following sequence of equations allowing one to determine 
the coefficient $\alpha$: 
\begin{eqnarray}
q^2=\alpha^2\,\tilde{q}^2 + 
\Bigg ( \sqrt{s_i\!-\!(\tilde{\bar{P}}\!-\!\frac{\alpha\tilde{q}}{2})^2} 
- \sqrt{s_f\!-\!(\tilde{\bar{P}}\!+\!\frac{\alpha\tilde{q}}{2})^2} \Bigg)^2\,,\hspace*{6cm}
\nonumber \\
\alpha^2\tilde{q}^2 
(4\bar{s}\!-\!q^2\!-\!4\tilde{\bar{P}}^2\!-\!4(\bar{P} \!\cdot\!\hat{\tilde{q}})^2)
\!+\!4\alpha \bar{P}\!\cdot\!\hat{\tilde{q}}\sqrt{-\tilde{q}^2}(s_i\!-\!s_f)
\!+\!(s_i\!-\!s_f)^2\!-\!q^2(4\bar{s}\!-\!q^2\!-\!4\tilde{\bar{P}}^2)=0\,.
\nonumber \\
\end{eqnarray}
The solution is given by:
\begin{eqnarray}
\alpha= \frac{\sqrt{-q^2}}{\sqrt{-\tilde{q}^2}}\,\frac{\sqrt{D0\;D2} \!+\!2
\bar{P}\!\cdot\!\hat{\tilde{q}}\,\frac{s_i\!-\!s_f}{Q}}{
(4\bar{s}\!-\!q^2\!-\!4\tilde{\bar{P}}^2\!-\!4(\bar{P}
\!\cdot\!\hat{\tilde{q}})^2)}=
\frac{\sqrt{-q^2}}{\sqrt{-\tilde{q}^2}}
\frac{D1}{\sqrt{D0\;D2} \!-\!2
\bar{P}\!\cdot\!\hat{\tilde{q}}\,\frac{s_i\!-\!s_f}{Q}}\,, \hspace*{2cm}
 \label{eq:if3}
\end{eqnarray}
where we used the following notations
\begin{eqnarray}
&&D0=4\bar{s} \!-\! q^2 \!-\! 4\tilde{\bar{P}}^2\, ,
\nonumber \\
&&D1=4\bar{s} \!-\! q^2\!+\! \frac{(s_i\!-\!s_f)^2}{Q^2}
  \!-\! 4\tilde{\bar{P}}^2\, ,
\nonumber \\
&&D2=4\bar{s} \!-\! q^2\!+\! \frac{(s_i\!-\!s_f)^2}{Q^2}
  \!-\! 4\tilde{\bar{P}}^2\!-\!4(\bar{P} \!\cdot\!\hat{\tilde{q}})^2\, .
 \label{eq:if4}
\end{eqnarray}
The above quantities are found to verify the following relations:
\begin{eqnarray}
&&D1^2=\Big (\sqrt{D0\;D2} \!-\!2
\bar{P}\!\cdot\!\hat{\tilde{q}}\,\frac{s_i\!-\!s_f}{Q}\Big)^2
+ \Big( 2\bar{P} \!\cdot\!\hat{\tilde{q}} \sqrt{D0}
 \!+\!\frac{s_i\!-\!s_f}{Q}\sqrt{D2}\Big)^2\, ,\hspace*{2cm}
\nonumber \\
&&D0\,D2-\Big (2
\bar{P}\!\cdot\!\hat{\tilde{q}}\,\frac{s_i\!-\!s_f}{Q}  \Big)^2
=D1\,(4\bar{s}\!-\!q^2\!-\!4\tilde{\bar{P}}^2\!-
\!4(\bar{P} \!\cdot\!\hat{\tilde{q}})^2)\,.\hspace*{2cm}
 \label{eq:if5}
\end{eqnarray}
It is noticed that $\alpha$ does not depend explicitly on the quantity 
$p \cdot\! \hat{\lambda}$, as well as on constituent masses $m_1,\;m_2$. 
The dependence that appeared in a previous work \cite{Desplanques:2008fg} 
is incorporated in the writing of $\alpha$ adopted in this work, 
through the dependence of the variables $s_{i,f}$ on these quantities. 

\noindent
{\bf Expression of $p \!\cdot\! \tilde{\bar{P}}$ and
 $p \!\cdot\! \hat{\tilde{q}}$ in terms of $s_i,\; s_f$ and 
$p\!\cdot\!\hat{\lambda}$ :}\\
From squaring Eq. (\ref{eq:if1}), we get  expressions of 
$p\!\cdot\!\tilde{P}_{i,f} $ in terms of the variables  $s_i,\; s_f$ 
and $p\!\cdot\! \hat{\lambda}$:
\begin{eqnarray}
2p \!\cdot\!   \tilde{P}_{i,f}=s^0_{i,f}\!+\!\Delta m^2
-2p  \!\cdot\! \hat{\lambda} \,\sqrt{s^0_{i,f}-\tilde{P}_{i,f} ^2 }\, .
 \label{eq:if6}
\end{eqnarray}
The appropriate expression that accounts for space-time translation properties
is given by:
\begin{eqnarray}
2p \!\cdot\! (  \tilde{\bar{P}} \mp \frac{\alpha\tilde{q}}{2})
=s_{i,f}\!+\!\Delta m^2
-2p  \!\cdot\! \hat{\lambda} \,
\sqrt{s_{i,f}\!-\!(\tilde{\bar{P}}\!\mp\!\frac{\alpha\tilde{q}}{2})^2 }\, .
 \label{eq:if7}
\end{eqnarray}
We now separate terms symmetrical and antisymmetrical in the exchange 
of initial and final states. In this order, we use the relation:
\begin{eqnarray}
\sqrt{s_{i,f}\!-\!(\tilde{\bar{P}}\!\mp\!\frac{\alpha\tilde{q}}{2})^2} 
=\frac{1}{2}\Bigg (\sqrt{D0}\pm Q \,
\frac{2 \bar{P} \!\cdot\!\hat{\tilde{q}} \sqrt{D0}
 \!+\!\frac{s_i\!-\!s_f}{Q}\sqrt{D2}}{\sqrt{D0\;D2} \!-\!
2 \bar{P} \!\cdot\!\hat{\tilde{q}}\,\frac{s_i\!-\!s_f}{Q}} \Bigg)\, ,
 \label{eq:if8}
\end{eqnarray}
and obtain:
\begin{eqnarray}
&&(p \!\cdot\!\tilde{\bar{P}})=
\frac{1}{2} \Big(\bar{s}\!+\!\Delta m^2-p \!\cdot\! \hat{\lambda}\,\sqrt{D0} \Big)\, ,
\nonumber \\
&&(p \!\cdot\!\hat{\tilde{q}})=\frac{1}{D1}
\Bigg (\!\!-\! \frac{s_i\!-\!s_f}{2Q} \Big (\sqrt{D0\;D2} \!-\!
2 \bar{P} \!\cdot\!\hat{\tilde{q}}\,\frac{s_i\!-\!s_f}{Q}\Big)
+ (p\!\cdot\! \hat{\lambda})
 \Big (2 \bar{P} \!\cdot\!\hat{\tilde{q}} \sqrt{D0}
 \!+\!\frac{s_i\!-\!s_f}{Q}\sqrt{D2} \Big)\Bigg)\, .
\nonumber \\
 \label{eq:if9}
\end{eqnarray}
A related but slightly simpler expression is :
\begin{eqnarray}
&&(\bar{P}\!-\!p) \!\cdot\! \hat{\tilde{q}}=
\frac{\sqrt{D0}\!-\!2p \!\cdot\!  \hat{\lambda}}{2D1}\; 
\Big(2\bar{P} \!\cdot\!\hat{\tilde{q}} \sqrt{D0}
 \!+\!\frac{s_i\!-\!s_f}{Q}\sqrt{D2}\Big) \, .
 \label{eq:if10}
\end{eqnarray}
Other useful relations are:
\begin{eqnarray}
``(p_i \!+\! p_f \!+\! 2p)" \!\cdot\!  \hat{\lambda}= \sqrt{D0}, \hspace*{1cm}
``(p_i \!+\! p_f )" \!\cdot\! \hat{\lambda}= 
    \sqrt{D0} \!-\! 2p \!\cdot\!  \hat{\lambda} \,.
 \label{eq:if11}
\end{eqnarray}
If necessary, one can invert the above expressions, 
Eqs. (\ref{eq:if9}), and get the quantities $s_i,\; s_f$ 
in terms of $p \!\cdot\!\tilde{\bar{P}}$ and $p \!\cdot\!\hat{\tilde{q}}$. 
This does not present major difficulty as the equations somewhat decouple.
The first one allows one to get $\bar{s}$ in terms 
of $p \!\cdot\!\tilde{\bar{P}}$. Using this result, the second equation can be
used to get $s_i- s_f$. The equation is in principle a 4rth degree one but,
depending on the square of the unknown quantity, it can be easily solved. 

\noindent
{\bf Jacobian}\\
The derivation of the Jacobian for the transformation of the momentum 
of the spectator constituent, $\vec{p}$, to the variables $s_i,\;s_f$ 
and $p\!\cdot\!  \hat{\lambda}$ can be performed in two steps: 
from the  $\vec{p}$ variable to $p \!\cdot\!\tilde{\bar{P}}$, 
$p \!\cdot\!\hat{\tilde{q}}$ and $p\!\cdot\!  \hat{\lambda}$
and from these ones to  $s_i,\;s_f$ and $p\!\cdot\!  \hat{\lambda}$. 
For the first step, one can  start from the following equation:
\begin{eqnarray}
 \frac{d\vec{p}}{e_p}=|J_1|\;d(p \!\cdot\!\tilde{\bar{P}})\, 
 d(p \!\cdot\!\hat{\tilde{q}}) \,d(p\!\cdot\!  \hat{\lambda})\, ,
 \label{eq:if12}
\end{eqnarray}
where $|J_1|$ is given in the present case by Eq. (112) of Ref. 
\cite{Desplanques:2008fg}: 
\begin{eqnarray}
|J_1|=\Big|(m_2^2 \!-\! (p\!\cdot\! \hat{\lambda})^2)
( \tilde{\bar{P}}^2\! +   \!  (\tilde{\bar{P}} \!\cdot\!\hat{\tilde{q}})^2 ) 
\!-\!2 (p \!\cdot\!\hat{\tilde{q}}) (p \!\cdot\!\tilde{\bar{P}}) 
(\tilde{\bar{P}} \!\cdot\!\hat{\tilde{q}})
\! -\!(p \!\cdot\!\tilde{\bar{P}})^2\!+\!\tilde{\bar{P}}^2  (p
\!\cdot\!\hat{\tilde{q}})^2\Big|^{-\frac{1}{2}}\,.
 \label{eq:if13}
\end{eqnarray}
The above quantity  can be expressed in terms of the variables $s_i,\;s_f$ 
and $p\!\cdot\!  \hat{\lambda}$ using Eqs. (\ref{eq:if9}). 
It may be written as:
\begin{eqnarray}
|J_1|=\frac{4D1}{\sqrt{D}\sqrt{D0}(\sqrt{D0\;D2} \!-\!
2 \bar{P} \!\cdot\!\hat{\tilde{q}}\,\frac{s_i\!-\!s_f}{Q})}
\frac{\theta(\cdots)}{\sqrt{\Big(\frac{s_is_f }{D}\,c_{\Delta m^2}\!-\!m_2^2\Big)f
- \Big(\frac{2p \cdot   \hat{\lambda}}{\sqrt{D0}}-d\Big)^2}}\, ,
 \label{eq:if14}
\end{eqnarray}
where $c_{\Delta m^2}$ has been defined in Eq. (\ref{eq:dp5}) 
and $\theta(\cdots)$ accounts for the fact that the last term makes sense 
only if the factor in front of the quantity $f$  is positive.
The quantities $d$ and $f$  are given by \footnote{The notation has been changed with
respect to the one used in  Ref. \cite{Desplanques:2008fg}. The motivation is to make the
presentation of present results closer to those obtained in the front form (see
next section)}:
\begin{eqnarray}
&&d=\frac{\frac{\sqrt{D2}}{D\sqrt{D0}}\Big (2(\bar{s}\!+\!\Delta m^2)D1\!-\!4\tilde{\bar{P}}^2\,
 \frac{(s_i\!-\!s_f)^2}{Q^2}\Big)
\!-\!2\bar{P}\!\cdot\!\hat{\tilde{q}}\,\frac{s_i\!-\!s_f}{Q}
}{\Big (\sqrt{D0\;D2} \!-\!2
\bar{P}\!\cdot\!\hat{\tilde{q}}\,\frac{s_i\!-\!s_f}{Q}\Big)}
\nonumber \\
&&\hspace*{1cm}=1\!-\! \frac{D1\sqrt{D2} }{\sqrt{D0}\Big (\sqrt{D0\;D2} \!-\!2
\bar{P}\!\cdot\!\hat{\tilde{q}}\,\frac{s_i\!-\!s_f}{Q}\Big)}
\frac{2\bar{s}\!-\! 2\Delta m^2 \!+\! Q^2}{D}
 \,,
\nonumber \\
&&f=
\frac{4D1^2( -4\tilde{\bar{P}}^2\! -   \!  4(\tilde{\bar{P}} \!\cdot\!\hat{\tilde{q}})^2 ) 
 }{D\,D0\,\Big (\sqrt{D0\;D2} \!-\!2
\bar{P}\!\cdot\!\hat{\tilde{q}}\,\frac{s_i\!-\!s_f}{Q}\Big)^2 }
=\frac{4}{D}\,\frac{ D1^2 (D2\!-\!D) }{D0\,\Big (\sqrt{D0\;D2} \!-\!2
\bar{P}\!\cdot\!\hat{\tilde{q}}\,\frac{s_i\!-\!s_f}{Q}\Big)^2 }\, .
 \label{eq:if15}
\end{eqnarray}
The second step is relatively easy as $p \!\cdot\!\tilde{\bar{P}}$ does not
depend on $(s_i\!-\!s_f)$. One thus has:
\begin{eqnarray}
&&\frac{d(p \!\cdot\!\tilde{\bar{P}})}{d\bar{s}}
=\frac{\sqrt{D0}\!-\!2p\!\cdot\!  \hat{\lambda}}{2\sqrt{D0}}\, ,
\nonumber \\
&&\frac{d(p \!\cdot\!\hat{\tilde{q}})}{d(s_i\!-\!s_f)}
=\frac{\sqrt{D0}\!-\!2p\!\cdot\!  \hat{\lambda}}{2QD1^2\sqrt{D2}}
\Big (\sqrt{D0\;D2} \!-\!
2 \bar{P} \!\cdot\!\hat{\tilde{q}}\,\frac{s_i\!-\!s_f}{Q}\Big)^2\, .
 \label{eq:if16}
\end{eqnarray}
The complete expression of the integration volume thus reads:
\begin{eqnarray}
\frac{d\vec{p}}{e_p} =
\sum \frac {d\bar{s} \; d(s_i\!-\!s_f)\,d(\frac{2p \cdot   \hat{\lambda}}{\sqrt{D0}}) \;
\theta(\cdots)\,\Big  (\sqrt{D0\;D2} \!-\!2
\bar{P}\!\cdot\!\hat{\tilde{q}}\,\frac{s_i\!-\!s_f}{Q}\Big)
}{2Q \,D1\,\sqrt{D2}\,\sqrt{D}\,\sqrt {\Big (\frac{s_is_f}{D}\,c_{\Delta m^2}\!-\!m_2^2  \Big )f \!-\! 
(\frac{2p \cdot   \hat{\lambda}}{\sqrt{D0}}\!-\!d)^2}}
\frac{(\sqrt{D0}\!-\!2p\!\cdot\!  \hat{\lambda})^2}{\sqrt{D0}}\, ,
 \label{eq:if17}
\end{eqnarray}
where the sum symbol accounts for the existence of two values of $p_z$ 
(and $e_p$) to be considered (see Sec. 2 and Eqs. (\ref{eq:dp2}, \ref{eq:dp3}) 
for a similar case).

\noindent
{\bf Charge form factor (scalar constituents)}\\
Results for the charge form factor at $Q^2=0$ are not affected 
by the implementation of constraints related
to space-time translation properties. To make these results
independent of the momentum of the system or of the front orientation 
(Lorentz invariance), 
a minimal factor has to be inserted in the integrand. This factor, given by 
$(``(2p\!+\!p_i\!+\!p_f)"\!\cdot\! \lambda) /
(2``(p_i\!+\!p_f)" \!\cdot\! \lambda) $,  can be seen to be equal to 
$\sqrt{D0}/(2(\sqrt{D0}\!-\!2p\!\cdot\!  \hat{\lambda}))$. Its
introduction removes one of the factors $(\sqrt{D0}\!-\!2p\!\cdot\! 
\hat{\lambda})$ at the r.h.s. of Eq. (\ref{eq:if17}). 
For the remaining factor, it is convenient to write it as follows: 
\begin{eqnarray}
\sqrt{D0}\!-\!2p \!\cdot\!   \hat{\lambda}=
\frac{D1\;\sqrt{D2}}{\Big (\sqrt{D0\;D2} \!-\!2
\bar{P}\!\cdot\!\hat{\tilde{q}}\,\frac{s_i\!-\!s_f}{Q}\Big) }
\frac{2\bar{s}\!-\! 2\Delta m^2 \!+\! Q^2}{D}
-\sqrt{D0}\,\Big(\frac{2p \!\cdot\!   \hat{\lambda}}{\sqrt{D0}}\!-\!d\Big)
\, ,
 \label{eq:if18}
\end{eqnarray}
where the second term at the r.h.s. gives zero upon integration 
over $p \cdot   \hat{\lambda}$. One thus gets:
\begin{eqnarray}
\frac{d\vec{p}}{e_p}\,\frac{``(2p\!+\!p_i\!+\!p_f)"\!\cdot\! \lambda 
}{2``(p_i\!+\!p_f)" \!\cdot\! \lambda} =
\sum \frac {d\bar{s} \; d(s_i\!-\!s_f)\,d(\frac{2p \cdot   \hat{\lambda}}{\sqrt{D0}}) \;
\theta(\cdots) \,\Big [\frac{(2\bar{s}\!-\! 2\Delta m^2 \!+\! Q^2)}{D}-
 (\frac{2p \cdot   \hat{\lambda}}{\sqrt{D0}}\!-\!d)g \Big]
}{4Q\,\sqrt{D} \sqrt {\Big (\frac{s_is_f}{D}\,c_{\Delta m^2}\!-\!m_2^2  \Big )f \!-\! 
(\frac{2p \cdot   \hat{\lambda}}{\sqrt{D0}}\!-\!d)^2}} \, ,
 \label{eq:if19}
\end{eqnarray}
where $g$ is given by:
\begin{eqnarray}
g=\frac { \Big (\sqrt{D0\;D2} \!-\!2
\bar{P}\!\cdot\!\hat{\tilde{q}}\,\frac{s_i\!-\!s_f}{Q}\Big)\sqrt{D0}  }{
D1\sqrt{D2}}
=\frac{D0}{D1}\Big (1\!-\!\frac{ 2
\bar{P}\!\cdot\!\hat{\tilde{q}}}{\sqrt{D0\;D2}}\,\frac{s_i\!-\!s_f}{Q} \Big)\,.
 \label{eq:if20}
\end{eqnarray}
We then obtain the following expression for the charge form factor:
\begin{eqnarray} 
``F_1(Q^2)"&\!=\!&\frac{16\pi^2}{N}\! \int\! \frac{d\vec{p}}{(2\pi)^3}\;
\frac{1}{e_p} \;^{``}\Bigg(
\frac {(p_i\!+\!p_f\!+\!2p)\!\cdot\!\lambda}{2(p_i\!+\!p_f)\!\cdot\!\lambda}   \;
\tilde{\phi}(\vec{k_f}^2)\;\tilde{\phi}(\vec{k_i}^2) \Bigg)^{"}
\nonumber \\
&\!=\!&\frac{2}{\pi N}\!\int \! \! \int d\bar{s} \; d(\frac{s_i\!-\!s_f}{Q}) 
\;\phi(s_f)\;\phi(s_i) \; \frac{ \theta(\cdots) }{4\sqrt{D}} \;
 \nonumber \\
&& \hspace*{0.6cm} \times
\sum \int \!\frac {d(\frac{2p \cdot   \hat{\lambda}}{\sqrt{D0}})  \;
\Big [\frac{2\bar{s}\!-\! 2\Delta m^2\!+\!Q^2 }{ D} 
\! -\!(\frac{2p \cdot   \hat{\lambda}}{\sqrt{D0}}\!-\!d)\,g \Big]   }{ 
\sqrt {\Big  (\frac {s_i\,s_f}{D }\,c_{\Delta m^2}\!-\!m^2\Big )f
\!-\! (  \frac{2p \cdot   \hat{\lambda}}{\sqrt{D0}}\!-\!d)^2}} 
 \nonumber \\
&\!=\!& \frac{1}{ N}\!\int \! \! \int d\bar{s} \; d(\frac{s_i\!-\!s_f}{Q})\; 
\phi(s_f)\;\phi(s_i)\;
\frac{ [2\bar{s}\!-\! 2\Delta m^2\!+\!Q^2] \; \theta(\cdots)  }{D\sqrt{D}} \, .
\label{eq:if21}
\end{eqnarray} 
The function $\theta(\cdots)$ has been defined in Eq. (\ref{eq:dp66}).
Performing the  operations at the third line in the above equation provides 
a factor $2\pi\frac{2\bar{s}\!-\! 2\Delta m^2\!+\!Q^2 }{ D}$, 
allowing one to recover  Eq. (\ref{eq:dp12}) for $F_1(Q^2)$.

\noindent
{\bf Charge form factor (spin-1/2 constituents)}\\
To account for the spin-1/2 nature of the constituents, one can introduce 
in  the integrand for the scalar-constituent case the ratio 
of the corresponding matrix elements of the current, 
$I_{\lambda}=I \cdot \lambda$ 
and $\tilde{I}_{\lambda} =\tilde{I}\cdot \lambda $:
\begin{eqnarray}
&& \frac{I_{\lambda}}{\tilde{I}_{\lambda}}=``\Bigg(
 \frac{(p_i\!+\!p_f)\!\cdot\! \hat{\lambda}\;\Big(\bar{s}^0\!-\!(m_1\!-\!m_2)^2\Big)
\!-\!(p_i\!-\!p_f)\!\cdot\! \hat{\lambda}\,\frac{s^0_i\!-\!s^0_f}{2} 
+p\!\cdot\!  \hat{\lambda}\,(p_i\!-\!p_f)^2
 }{(p_i\!+\!p_f)\!\cdot\! \hat{\lambda}
\;\sqrt{s^0_i\!-\!(m_1\!-\!m_2)^2}\;\sqrt{s^0_f\!-\!(m_1\!-\!m_2)^2}}\Bigg)"
 \nonumber \\
&& \hspace*{0.8cm}=
 \frac{(\sqrt{D0}\!-\!2p\!\cdot\!  \hat{\lambda})\Big(\bar{s}\!-\!(m_1\!-\!m_2)^2\Big)
-Q\,\frac{s_i\!-\!s_f}{2} 
\frac{2 \bar{P} \!\cdot\!\hat{\tilde{q}} \sqrt{D0}
 \!+\!\frac{s_i\!-\!s_f}{Q}\sqrt{D2}}{\sqrt{D0\;D2} \!-\!
2 \bar{P} \!\cdot\!\hat{\tilde{q}}\,\frac{s_i\!-\!s_f}{Q}}
 +p\!\cdot\!  \hat{\lambda}\,q^2
 }{(\sqrt{D0}\!-\!2p\!\cdot\!  \hat{\lambda}) \,
\sqrt{s_i\!-\!(m_1\!-\!m_2)^2}\;\sqrt{s_f\!-\!(m_1\!-\!m_2)^2}}\,,
\label{eq:if22}
\end{eqnarray}
where the numerator, similarly to Eq. (\ref{eq:if18}), can be written as the
sum of a term independent of $p\!\cdot\!  \hat{\lambda}$ and another one 
that will give 0 upon integration on this variable:
\begin{eqnarray}
&&(\sqrt{D0}\!-\!2p\!\cdot\!  \hat{\lambda})\Big(\bar{s}\!-\!(m_1\!-\!m_2)^2\Big)
-Q\,\frac{s_i\!-\!s_f}{2} 
\frac{2 \bar{P} \!\cdot\!\hat{\tilde{q}} \sqrt{D0}
 \!+\!\frac{s_i\!-\!s_f}{Q}\sqrt{D2}}{\sqrt{D0\;D2} \!-\!
2 \bar{P} \!\cdot\!\hat{\tilde{q}}\,\frac{s_i\!-\!s_f}{Q}} +p\!\cdot\!  \hat{\lambda}\,q^2
 \nonumber \\
&&
= \frac{D1\;\sqrt{D2}}{\Big (\sqrt{D0\;D2} \!-\!2
\bar{P}\!\cdot\!\hat{\tilde{q}}\,\frac{s_i\!-\!s_f}{Q}\Big) }
\frac{\Big [2s_is_f\!-\!\Delta m^2(2\bar{s}\!+\!Q^2) 
\!-\!(m_1\!-\!m_2)^2 (2\bar{s}\!-\!2\Delta m^2\!+\!Q^2)\Big]}{D}
 \nonumber \\
&&\hspace*{3cm}-\sqrt{D0}\,\Big(  \frac{2p \cdot   \hat{\lambda}}{\sqrt{D0}}
\!-\!d\Big)\,\Big[\bar{s}\!-\!(m_1\!-\!m_2)^2\!+\!\frac{Q^2}{2}\Big] \, .
\end{eqnarray}
Introducing the above ratios in Eqs. (\ref{eq:if21}), one gets the expression
of the charge form factor for  spin-1/2 constituents:
\begin{eqnarray} 
``F_1(Q^2)"&\!=\!&\frac{2}{\pi N}\!\int \! \! \int d\bar{s} \; d(\frac{s_i\!-\!s_f}{Q}) 
\frac{\phi(s_f)\;\phi(s_i) \;\theta(\cdots)}{4\sqrt{D}\;\sqrt{s_i\!-\!(m_1\!-\!m_2)^2}\;\sqrt{s_f\!-\!(m_1\!-\!m_2)^2}} 
\nonumber \\
&& \hspace*{0.6cm} \times
\sum \int \!\frac {d(\frac{2p \cdot   \hat{\lambda}}{\sqrt{D0}})  \;
\Big [\frac{2s_is_f\!-\!\Delta m^2(2\bar{s}\!+\!Q^2) 
\!-\!(m_1\!-\!m_2)^2 (2\bar{s}\!-\!2\Delta m^2\!+\!Q^2)}{ D} 
\! -\!(\frac{2p \cdot   \hat{\lambda}}{\sqrt{D0}}\!-\!d)\,g' \Big]   }{ 
\sqrt {\Big  (\frac {s_i\,s_f}{D }\,c_{\Delta m^2}\!-\!m^2\Big )f
\!-\! (  \frac{2p \cdot   \hat{\lambda}}{\sqrt{D0}}\!-\!d)^2}} 
 \nonumber \\
 & \!=\! & \frac{1}{ N} \int \!  d\bar{s} \;  d(\frac{s_i\!-\!s_f}{Q})\;
 \phi(s_i) \; \phi(s_f)
\nonumber \\
&& \times
 \frac{ \Big [2s_is_f\!-\!\Delta m^2(2\bar{s}\!+\!Q^2) 
\!-\!(m_1\!-\!m_2)^2 (2\bar{s}\!-\!2\Delta m^2\!+\!Q^2)\Big]\;\theta(\cdots)}{
D\sqrt{D}\;\sqrt{s_i\!-\!(m_1\!-\!m_2)^2}\;\sqrt{s_f\!-\!(m_1\!-\!m_2)^2} }
 \;,
\label{eq:if23}
\end{eqnarray} 
where the quantity $g'$ is given by:
\begin{eqnarray}
g'=g\;\Big[\bar{s}\!-\!(m_1\!-\!m_2)^2\!+\!\frac{Q^2}{2}\Big]\, . 
\label{eq:if24}
\end{eqnarray} 
%

\section{Incorporation of different masses (and spin 1/2) 
in the scalar-constituent results: generalized front form} 
The relation to start with in the generalized front-form case
with orientation  $\omega^{\mu}$ ($\omega^2=0$)  is given by Eq. (16) of Ref. \cite{Desplanques:2008fg}:
\begin{eqnarray} 
p_{i,f}^{\mu}&\!=\!&P_{i,f}^{\mu}-p^{\mu}
+\omega^{\mu}\;\frac{(s^0_{i,f}\!-\!P_{i,f}^2)}{2 P_{i,f}\! \cdot\! \omega}\, .
\label{eq:ff1}
\end{eqnarray}
{\bf Derivation of $\alpha$:}\\
One has the following set of equations:
\begin{eqnarray} 
p_i^{\mu}-p_f^{\mu}&\!=\!&P_i^{\mu}\!-\!P_f^{\mu}
+\omega^{\mu}\,\Bigg (\frac{s^0_i\!-\!P_i^2}{2 P_i \!\cdot\! \omega}
                  - \frac{s^0_f\!-\!P_f^2}{2 P_f \!\cdot\! \omega} \Bigg)\, ,
\nonumber \\
(p_i\!-\!p_f)^2&\!=\!& (P_i\!-\!P_f)^2
+2 (P_i\!-\!P_f) \!\cdot\! \omega
   \,\Bigg (\frac{s^0_i\!-\!P_i^2}{2 P_i \!\cdot\! \omega}
         - \frac{s^0_f\!-\!P_f^2}{2 P_f \!\cdot\! \omega} \Bigg)^2\, .
\label{eq:ff2}
\end{eqnarray}
Accounting for corrections related to space-time translation properties, 
one gets the following sequence of equations allowing one to obtain 
the coefficient $\alpha$:
\begin{eqnarray}
&&q^2=\alpha^2\,q^2 -2\alpha \, q \!\cdot\! \omega \,
\Bigg (\frac{s_i\!-\!(\bar{P}\!-\! \frac{\alpha q}{2})^2  }{ 
           2(\bar{P}\!-\! \frac{\alpha q}{2}) \!\cdot\! \omega}
    -  \frac{s_f\!-\!(\bar{P}\!+\!\frac{\alpha q}{2})^2  }{
           2(\bar{P}\!+\! \frac{\alpha q}{2}) \!\cdot\! \omega} \Bigg)^2\, ,
\nonumber \\
&&\alpha^2\Big(-(\bar{P} \!\cdot\! \omega)^2  q^2
+2q \!\cdot\! \omega \, \bar{P} \!\cdot\! \omega \, \bar{P} \!\cdot\! q 
+(q \!\cdot\! \omega)^2) (\bar{s}-\bar{P}^2-\frac{q^2}{4})
\Big)
\nonumber \\
&&\hspace*{2cm}
+\alpha  \,q \!\cdot\! \omega \,  \bar{P} \!\cdot\! \omega (s_i\!-\!s_f)
+q^2( \bar{P} \!\cdot\! \omega )^2=0 \,.
\end{eqnarray}
The solution is given by:
\begin{eqnarray}
&&\alpha=\frac{1}{\sqrt{E2} -
\frac{q\cdot \omega }{ \bar{P}\cdot\omega }\,\frac{s_i\!-\!s_f}{2q^2}}\,,
\label{eq:ff3}
\end{eqnarray}
where $E2$ is given by: 
\begin{eqnarray}
E2&=&1
\!-\!2\frac{q \!\cdot\! \omega }{ \bar{P} \!\cdot\! \omega } 
\frac{\bar{P} \!\cdot\! q}{q^2}
\!+\!\Big(\frac{q \!\cdot\! \omega }{ \bar{P} \!\cdot\! \omega }\Big)^2\;
\frac{1 }{q^2} \Big( \bar{P}^2 \!+\! \frac{q^2}{4}\!-\!\bar{s}
\!+\! \frac{(s_i\!-\!s_f)^2}{4q^2}\Big) \, .
\nonumber \\ 
\label{eq:ff4}
\end{eqnarray}
The factor $\alpha$ does not depend explicitly on  $p \!\cdot\! \omega$, 
as well as on constituent masses $m_1,\;m_2$.\\

\noindent
{\bf Expressions of $p \!\cdot\! \bar{P}$ and  $p \!\cdot\! q$ 
in terms of $s_i, s_f$ and $q \!\cdot\! \omega$}\\
From squaring Eq. (\ref{eq:ff1}), one  obtains:
\begin{eqnarray}
2p \!\cdot\!   P_{i,f}=s^0_{i,f}\!+\!\Delta m^2
-\frac{p  \!\cdot\! \omega }{P_{i,f}\!\cdot\! \omega}\;(s^0_{i,f}-P_{i,f} ^2 )
\label{eq:ff5}
\end{eqnarray}
Accounting for effects related to space-time translation properties, one gets:
\begin{eqnarray}
2p \!\cdot\!   (\bar{P}\!\mp\! \frac{\alpha q}{2})=
s_{i,f}\!+\!\Delta m^2
-\frac{p  \!\cdot\! \omega }{
(\bar{P}\!\mp\! \frac{\alpha q}{2})\!\cdot\! \omega}\;
\Big(s_{i,f}-(\bar{P}\!\mp\! \frac{\alpha q}{2}) ^2 \Big)\,.
\label{eq:ff6}
\end{eqnarray}
We can now separate the parts symmetrical and antisymmetrical in the initial and
final states. In this order, we use the following relation: 
\begin{eqnarray}
\frac{ s_{i,f}\!-\!(\bar{P}\!\mp\! \frac{\alpha\, q}{2}) ^2 }{
(1\!\mp\! \frac{\alpha \,q \cdot  \omega}{2 \bar{P} \cdot  \omega})}
=\bar{s} \!-\! \bar{P}^2 \!-\! \frac{q^2}{4}
\pm \Bigg (\frac{s_i\!-\!s_f}{2}  \!+\! 
\frac{\bar{P} \!\cdot\! q \!-\!
\frac{q  \cdot  \omega }{ 2 \bar{P} \cdot  \omega }
 \Big( \bar{P}^2 \!+\! \frac{q^2}{4}\!-\!\bar{s}\Big)
}{ \sqrt{E2} -
  \frac{q  \cdot  \omega }{ \bar{P}  \cdot  \omega }
  \, \frac{s_i\!-\!s_f}{2q^2}}  \Bigg)\, ,
\label{eq:ff7}
\end{eqnarray}
and obtain:
\begin{eqnarray}
&&p\!\cdot\!\bar{P}=\frac{1}{2}\Bigg(\frac{p\!\cdot\! \omega }{ \bar{P}\!\cdot\!\omega }
\Big(\bar{P}^2\!+\!\frac{q^2}{4}\!-\!\bar{s}\Big)
 \!+\!  \bar{s} \!+\! \Delta m^2\Bigg)\,,
\nonumber \\
&&p\!\cdot\!q=-\frac{s_i\!-\!s_f}{2} \Bigg(\sqrt{E2} -
\frac{q \!\cdot\! \omega }{ \bar{P} \!\cdot\! \omega }
\,\frac{s_i\!-\!s_f}{2q^2}\Bigg)
\frac{(\bar{P}\!-\!p)\!\cdot\! \omega }{ \bar{P}\!\cdot\!\omega }
\nonumber \\
&&\hspace*{3cm}+\frac{p\!\cdot\! \omega }{ \bar{P}\!\cdot\!\omega }
\Bigg(  \bar{P}\!\cdot\!q\!-\!\frac{1}{2}
\frac {q\!\cdot\!\omega}{\bar{P}\!\cdot\! \omega}
\Big (\bar{P}^2 \!+\! \frac{q^2}{4}\!-\!\bar{s} \Big)  \Bigg)\, .
\label{eq:ff8}
\end{eqnarray}
Related equations are:
\begin{eqnarray}
&&\bar{P}^2\!-\!2p \!\cdot\! \bar{P}\!+\!\frac{q^2}{4}\!+\!\Delta m^2=
\frac{(\bar{P}\!-\!p)\!\cdot\! \omega }{ \bar{P}\!\cdot\!\omega }
\Big(\bar{P}^2\!+\!\frac{q^2}{4}\!-\!\bar{s}\Big)\, ,
\nonumber \\
&&p\!\cdot\!q=-\frac{s_i\!-\!s_f}{2} \Bigg(\!\sqrt{E2} -
\frac{q \!\cdot\! \omega }{ \bar{P} \!\cdot\! \omega }
\,\frac{s_i\!-\!s_f}{2q^2}\Bigg)
\!-\! \frac{ p \!\cdot\! \omega\;  q^2}{2q \!\cdot\! \omega }
\Bigg (\Big (\!\sqrt{E2} -
  \frac{q \!\cdot\! \omega }{ \bar{P} \!\cdot\! \omega }
  \, \frac{s_i\!-\!s_f}{2q^2}\Big)^2 \!-\!1\Bigg)\, .
\nonumber \\
\label{eq:ff9}
\end{eqnarray}
Inverting  Eqs. (\ref{eq:ff8}) to get $s_i,\;s_f$ in terms of $p\!\cdot\!\bar{P}$
and $p\!\cdot\!q$ can be done similarly to the generalized hyperplane case
discussed in the previous section.

\noindent
{\bf Jacobian}\\
The derivation of the Jacobian for the transformation of the momentum 
of the spectator constituent, $\vec{p}$, to the variables $s_i,\;s_f$ 
and $p\!\cdot\!  \omega$ can be performed in two steps: 
from the  $\vec{p}$ variable to $p \!\cdot\! \bar{P}$, 
$p \!\cdot\!q$ and $p\!\cdot\!  \omega$
and from these ones to  $s_i,\;s_f$ and $p\!\cdot\!  \omega$. 
For the first step, we  start from an equation similar to Eq. (\ref{eq:if12})  
for the generalized instant form:  
\begin{eqnarray}
 \frac{d\vec{p}}{e_p}=|J_1|\;d(p \!\cdot\! \bar{P})\, 
 d(p \!\cdot\! q) \,d(p\!\cdot\!  \omega)\, .
\label{eq:ff10}
\end{eqnarray}
The quantity $|J_1|$ can be calculated from Eqs. (106) of Ref. 
\cite{Desplanques:2008fg}
with the 4-vectors $a$, $b$ and $c$ replaced by $\omega$, $\bar{P}$  
and $q$ respectively. Reminding that $\omega^2=0$, one gets: 
\begin{eqnarray}
|J_1|\!&\!=\!&\!\Bigg|(\bar{P}\!\cdot\!\omega)^2 \Bigg(  m_2^2\Big (q^2
\!-\!2\frac{q\!\cdot\! \omega }{ \bar{P}\!\cdot\!\omega }\bar{P}\!\cdot\!q
\!+\!\Big (\frac{q\!\cdot\! \omega }{ \bar{P}\!\cdot\!\omega }\Big)^2
\bar{P}^2 \Big)
-\Big (p\!\cdot\!q- 
\frac{q\!\cdot\! \omega }{ \bar{P}\!\cdot\!\omega } \,
p\!\cdot\! \bar{P}\Big)^2
\nonumber \\
&&+2\frac{p\!\cdot\! \omega }{ \bar{P}\!\cdot\!\omega }
\Big (p\!\cdot\!q  (\bar{P}\!\cdot\!q
\!-\!\frac{q\!\cdot\! \omega }{ \bar{P}\!\cdot\!\omega }\bar{P}^2)
\!- \!p\!\cdot\! \bar{P} (q^2
\!-\!\frac{q\!\cdot\! \omega }{ \bar{P}\!\cdot\!\omega }
\bar{P}\!\cdot\!q )\Big)
+\Big (\frac{p\!\cdot\! \omega }{ \bar{P}\!\cdot\!\omega }\Big)^2
\Big (\bar{P}^2\,q^2\!-\!(\bar{P}\!\cdot\!q)^2\Big)\Bigg )\Bigg|^{-\frac{1}{2}}\, .
\nonumber \\ 
\label{eq:ff11}
\end{eqnarray}
Taking into account the expressions of $p\!\cdot\!\bar{P}$, $p\!\cdot\!q$ 
in terms of $s_i$, $s_f$ and $p\!\cdot\!\omega$, one also gets (after
rearranging the various terms):
\begin{eqnarray}
|J_1|\!&\!=\!&\!\frac{2}{\bar{P} \!\cdot\! \omega}\,
\frac{1}{ Q\sqrt{D} \Big (\sqrt {E2}-
\frac{q \cdot \omega }{ \bar{P} \cdot \omega } 
\,\frac{s_i\!-\!s_f}{2q^2} \Big )}
\,\frac{\theta(\cdots)}{
\sqrt{\Big  (\frac {s_i\,s_f}{D }\,c_{\Delta m^2}-m_2^2\Big )f 
-(\frac{p \cdot \omega}{\bar{P} \cdot \omega}\!-\!d)^2} }\, ,
\label{eq:ff12}
\end{eqnarray}
where $c_{\Delta m^2}$ has already been defined and $\theta(\cdots)$ accounts
for the fact that the factor in front of $f$ under the square-root symbol 
should be positive. The quantities $d$ and $f$ are now given by:
\begin{eqnarray}
&&d=\frac { (2\bar{s}+\!2\Delta m^2\!-\!\frac{(s_i\!-\!s_f)^2}{q^2}) \sqrt {E2}
 - \frac {q \cdot  \omega }{ \bar{P} \cdot \omega } 
\, \frac {s_i\!-\!s_f}{2q^2} D }{
D\Big (\sqrt {E2}- \frac{q \cdot  \omega }{ \bar{P} \cdot \omega } 
\frac{s_i\!-\!s_f}{2q^2}\Big ) }
=1-\frac { (2\bar{s}\!-\!2\Delta m^2\!-\!q^2) \sqrt {E2} }{
D \Big (\sqrt {E2}\!-\! \frac{q \cdot \omega }{ \bar{P} \cdot \omega } 
\frac{s_i\!-\!s_f}{2q^2}\Big ) }\, ,
\nonumber \\ 
&&f=\frac{4}{D}\,\frac {\Big (1 \!+\! 2\frac{\hat{q} \cdot  \omega }{ 
\bar{P} \cdot \omega }\bar{P} \!\cdot\! \hat{q}
\!-\! \Big (\frac{\hat{q} \cdot  \omega }{ \bar{P} \cdot \omega }\Big)^2
\bar{P}^2 \Big)}{ \Big (\!\sqrt {E2}\!-\!
\frac{q \cdot  \omega }{ \bar{P} \cdot \omega } 
\frac{s_i\!-\!s_f}{2q^2}
\Big )^2}\, .
\label{eq:ff13}
\end{eqnarray}
The second step involves the derivatives of $p\!\cdot\!\bar{P}$ with respect 
to $\frac{(s_i\!+\!s_f)}{2}$ and  $p\!\cdot\!q$ with respect 
to $(s_i\!-\!s_f)$. They are given by: 
\begin{eqnarray}
&&d(p\!\cdot\!\bar{P})=d \frac{(s_i\!+\!s_f)}{2}\times 
\frac{1}{2}\frac{(\bar{P}\!-\!p)\!\cdot\! \omega }{ \bar{P}\!\cdot\!\omega }\,,
\nonumber \\ 
&&d(p\!\cdot\!q)=-d(s_i\!-\!s_f)\times
\frac{(\bar{P}\!-\!p)\!\cdot\! \omega }{ \bar{P}\!\cdot\!\omega }
 \frac {1}{2\sqrt {E2}}
 \Bigg (\sqrt {E2}- \frac{q\!\cdot\! \omega }{ \bar{P} \!\cdot\! \omega } 
\,\frac{s_i\!-\!s_f}{2q^2} \Bigg )^2\,.
\label{eq:ff14}
\end{eqnarray}
Putting all factors together, we now get:
\begin{eqnarray}
&&\frac{d\vec{p}}{e_p} \; 
=\sum\frac{1}{2}\frac {d\bar{s}\, d(s_i\!-\!s_f)\, 
d(\frac {p \cdot \omega}{\bar{P} \cdot \omega })\, \theta(\cdots)
\;\Big (\frac {(\bar{P}\!-\!p) \cdot  \omega }{ \bar{P} \cdot \omega }\Big)^2 
\;\Big (\sqrt {E2}\!-\! \frac{q \cdot  \omega }{ \bar{P}  \cdot  \omega } 
\,\frac{s_i\!-\!s_f}{2q^2} \Big )}{  
Q\sqrt{ D}\,\sqrt{E2}\,\sqrt{\Big  (\frac {s_i\,s_f}{D }\,c_{\Delta m^2}
-m_2^2\Big )f 
-(\frac {p \cdot \omega}{\bar{P} \cdot \omega}\!-\!d)^2} }\, ,
\label{eq:ff15}
\end{eqnarray}
where the sum symbol accounts for the fact that there are two acceptable values of
$p_z$ (and $e_p$) corresponding to the same set of variables $\bar{s}$, 
$s_i-s_f$ and $\frac{p \cdot \omega}{\bar{P} \cdot \omega}$.

\noindent
{\bf Charge form factor (scalar constituents)}\\
Results for the charge form factor at $Q^2=0$ are not affected 
by the implementation of constraints related
to space-time translation properties. To make these results
independent of the momentum of the system or of the front orientation
(Lorentz invariance), 
a minimal factor has to be inserted in the integrand. This factor, given by 
$(``(2p\!+\!p_i\!+\!p_f)"\!\cdot\! \omega) /
(2``(p_i\!+\!p_f)" \!\cdot\! \omega) $,  can be seen to be equal to 
$2\bar{P} \cdot \omega / ((\bar{P}\!-\!p) \cdot  \omega)$. Its
introduction removes one of the factors 
$ ((\bar{P}\!-\!p) \cdot  \omega) / \bar{P} \cdot \omega $ 
at the r.h.s. of Eq. (\ref{eq:ff15}). 
For the remaining factor, it is convenient to write it as follows: 
\begin{eqnarray}
\frac{(\bar{P}\!-\!p) \!\cdot\!  \omega}{\bar{P} \!\cdot\! \omega }=
\frac { (2\bar{s}\!-\!2\Delta m^2\!-\!q^2) \sqrt {E2} }{
D \Big (\sqrt {E2}\!-\! \frac{q \cdot \omega }{ \bar{P} \cdot \omega } 
\frac{s_i\!-\!s_f}{2q^2}\Big ) }
 -(\frac{p \!\cdot\! \omega}{\bar{P} \!\cdot\! \omega}\!-\!d)
\label{eq:ff16}
\end{eqnarray}
One thus gets:
\begin{eqnarray}
\frac{d\vec{p}}{e_p} \; 
\frac{``(p_i\!+\!p_f\!+\!2p)"\!\cdot\!\omega}{2``(p_i\!+\!p_f)"\!\cdot\!\omega} \;
\!&\!=\!&\sum \frac{d\bar{s}\, d(s_i\!-\!s_f)\, 
d(\frac{p \cdot \omega}{\bar{P} \cdot \omega })\,\theta(\cdots) 
\;\Big ( \frac{2\bar{s}\!-\!2\Delta m^2\!+\!Q^2 }{  D} 
 -(\frac{p \cdot \omega}{\bar{P} \cdot \omega}\!-\!d) g \Big) }{  
4Q\sqrt{ D}\sqrt{\Big  (\frac {s_i\,s_f}{D }\,c_{\Delta m^2}-m_2^2\Big )f 
-(\frac{p \cdot \omega}{\bar{P} \cdot \omega}\!-\!d)^2} } \, ,
\label{eq:ff17}
\end{eqnarray}
where $g$ is now given by: 
\begin{eqnarray}
g=1+ \frac{\hat{q} \ccdot \omega }{ \bar{P} \ccdot \omega } \,
\frac{s_i\!-\!s_f}{2Q\sqrt {E2}}\,.
\label{eq:ff18}
\end{eqnarray}
The charge form factor for scalar constituents then reads:
\begin{eqnarray} 
``F_1(Q^2)"&\!=\!&\frac{16\pi^2}{N}\! \int\! \frac{d\vec{p}}{(2\pi)^3}\;
\frac{1}{e_p} \;^{``}\Bigg(
\frac {(p_i\!+\!p_f\!+\!2p)\!\cdot\!\omega}{2(p_i\!+\!p_f)\!\cdot\!\omega}   \;
\tilde{\phi}(\vec{k_f}^2)\;\tilde{\phi}(\vec{k_i}^2) \Bigg)^{"}
\nonumber \\
&\!=\!&\frac{2}{\pi N}\!\int \! \! \int d\bar{s} \; d(\frac{s_i\!-\!s_f}{Q}) \;
\phi(s_f)\;\phi(s_i) \; \frac{ \theta(\cdots) }{4\sqrt{D}} \;
 \nonumber \\
&& \hspace*{0.6cm} \times
\sum \int \!\frac {d(\frac{p \cdot \omega}{\bar{P} \cdot \omega }) \; 
\Big [\frac{2\bar{s}\!-\! 2\Delta m^2\!+\!Q^2 }{ D} 
 -(\frac{p \cdot \omega}{\bar{P} \cdot \omega}-d)\,g \Big]   }{ 
\sqrt {\Big  (\frac {s_i\,s_f}{D }\,c_{\Delta m^2}-m_2^2\Big )f
-(\frac{p \cdot \omega}{\bar{P} \cdot \omega }\!-\!d)^2}}  
\nonumber \\
&\!=\!&\frac{1}{N}\!\int \! \! \int d\bar{s} \; d(\frac{s_i\!-\!s_f}{Q}) \;
\phi(s_f)\;\phi(s_i) \;
\frac{ (2\bar{s}\!-\! 2\Delta m^2\!+\!Q^2 )\,\theta(\cdots)  }{D\sqrt{D}} \, .
\label{eq:ff19}
\end{eqnarray} 

\noindent
{\bf Charge form factor (spin-1/2 constituents)}\\
To account for the spin-1/2 nature of the constituents, one can proceed 
in a way similar to the previous generalized instant form. One introduces 
in  the integrand for the scalar-constituent case the ratio 
of the corresponding matrix elements of the current, 
 $I_{\omega} =I \cdot \omega$ 
and $ \tilde{I}_{\omega}=\tilde{I} \cdot \omega $:
\begin{eqnarray}
&& \frac{I_{\omega}}{\tilde{I}_{\omega}}=``\Bigg(
 \frac{(p_i\!+\!p_f)\!\cdot\! \omega\;\Big(\bar{s}^0\!-\!(m_1\!-\!m_2)^2\Big)
 \!-\!(p_i\!-\!p_f)\!\cdot\! \omega\,\frac{s^0_i\!-\!s^0_f}{2}
+p\!\cdot\!  \omega\,(p_i\!-\!p_f)^2}{(p_i\!+\!p_f)\!\cdot\! \omega
\;\sqrt{s^0_i\!-\!(m_1\!-\!m_2)^2}\;\sqrt{s^0_f\!-\!(m_1\!-\!m_2)^2}}\Bigg)"
 \nonumber \\
&& \hspace*{0.8cm}=
 \frac{2(\bar{P}\!-\!p) \!\cdot\!  \omega \;\Big(\bar{s}\!-\!(m_1\!-\!m_2)^2\Big)
 +q\!\cdot\!  \omega  \,\frac{s_i\!-\!s_f}{2} 
\frac{1}{\sqrt {E2}\!-\! \frac{q \cdot \omega }{ \bar{P} \cdot \omega } 
\frac{s_i\!-\!s_f}{2q^2}} +p\!\cdot\!  \omega\,q^2
}{2(\bar{P}\!-\!p) \!\cdot\!  \omega  \,\sqrt{s_i\!-\!(m_1\!-\!m_2)^2}\;
\sqrt{s_f\!-\!(m_1\!-\!m_2)^2}}\,,
\label{eq:ff20}
\end{eqnarray}
where the numerator, similarly to Eq. (\ref{eq:ff16}), can be written as the
sum of a term independent of $p\!\cdot\!  \omega$ and another one 
that will give 0 upon integration on this variable:
\begin{eqnarray}
&&2(\bar{P}\!-\!p) \!\cdot\!  \omega \;\Big(\bar{s}\!-\!(m_1\!-\!m_2)^2\Big)
 +q\!\cdot\!  \omega  \,\frac{s_i\!-\!s_f}{2} 
\frac{1}{\sqrt {E2}\!-\! \frac{q \cdot \omega }{ \bar{P} \cdot \omega } 
\frac{s_i\!-\!s_f}{2q^2}} +p\!\cdot\!  \omega\,q^2
 \nonumber \\
&& \hspace*{0.8cm}=
2 \frac{\bar{P} \ccdot \omega \;\sqrt{E2}}{ \sqrt {E2}\!-\! \frac{q \cdot \omega }{ \bar{P} \cdot \omega } 
\frac{s_i\!-\!s_f}{2q^2}}
\frac{\Big [2s_is_f\!-\!\Delta m^2(2\bar{s}\!+\!Q^2) 
\!-\!(m_1\!-\!m_2)^2 (2\bar{s}\!-\!2\Delta m^2\!+\!Q^2)\Big]}{D}
 \nonumber \\
&&\hspace*{3cm}-2\bar{P} \ccdot \omega\,
\Big(  \frac{p \ccdot \omega}{\bar{P} \ccdot \omega }
\!-\!d\Big)\,\Big[\bar{s}\!-\!(m_1\!-\!m_2)^2\!+\!\frac{Q^2}{2}\Big] \, .
\label{eq:ff21}
\end{eqnarray}
Introducing the above ratios in Eq. (\ref{eq:ff19}), one gets the expression
of the charge form factor for spin-1/2 constituents:
\begin{eqnarray} 
``F_1(Q^2)"&\!=\!&\frac{2}{\pi N}\!\int \! \! \int d\bar{s} \; 
d(\frac{s_i\!-\!s_f}{Q}) \;
   \frac{  \phi(s_f)\;\phi(s_i) \;\theta(\cdots) }{   4\sqrt{D}\;
\sqrt{s_i\!-\!(m_1\!-\!m_2)^2}\;\sqrt{s_f\!-\!(m_1\!-\!m_2)^2}} 
\nonumber \\
&& \hspace*{0.6cm} \times
\sum \int \!\frac { d(\frac{p \cdot \omega}{\bar{P} \cdot \omega }) \; 
\Big [\frac{2s_is_f\!-\!\Delta m^2(2\bar{s}\!+\!Q^2) 
\!-\!(m_1\!-\!m_2)^2 (2\bar{s}\!-\!2\Delta m^2\!+\!Q^2)}{ D} 
 -(\frac{p \cdot \omega}{\bar{P} \cdot \omega}-d)\,g' \Big]}{ 
\sqrt {\Big  (\frac {s_i\,s_f}{D }\,c_{\Delta m^2}-m_2^2\Big )f
-(\frac{p \cdot \omega}{\bar{P} \cdot \omega }\!-\!d)^2}}  
\nonumber \\
&\!=\!&\frac{1}{N}\!\int \! \! \int d\bar{s} \; d(\frac{s_i\!-\!s_f}{Q}) \;
\phi(s_f)\;\phi(s_i)
\nonumber \\
&& \times
 \frac{ \Big [2s_is_f\!-\!\Delta m^2(2\bar{s}\!+\!Q^2) 
\!-\!(m_1\!-\!m_2)^2 (2\bar{s}\!-\!2\Delta m^2\!+\!Q^2)\Big]\;\theta(\cdots)}{
D\sqrt{D}\;\sqrt{s_i\!-\!(m_1\!-\!m_2)^2}\;\sqrt{s_f\!-\!(m_1\!-\!m_2)^2} } \, ,
\label{eq:ff22}
\end{eqnarray} 
where $g'$ is given by:
\begin{eqnarray}
g'=g\;\Big[\bar{s}\!-\!(m_1\!-\!m_2)^2\!+\!\frac{Q^2}{2}\Big]\, . 
\label{eq:ff23}
\end{eqnarray} 

\noindent
{\bf Relationship to other approaches}\\
Though results for various approaches were obtained independently, 
there are relations between those presented in this section and those presented
in previous ones for the front form with $q^+=0$ and the generalized instant
form for an arbitrary front orientation. The comparison is facilitated by the
conventions adopted in this paper. Thus, front-form results with $q^+=0$ 
can be obtained from those given in this section by assuming 
$q \cdot \omega = 0$ and writing $\frac{p \cdot \omega}{\bar{P} \cdot \omega }=x$. 
On the other hand, present front-form results can be obtained 
from the generalized instant-form ones by taking the limit 
$\omega^2 \rightarrow 0$, which could require some care. A few relations that
are helpful are the following ones:
\begin{eqnarray}
&&\Big(\frac{D0}{D1}\Big)_{\omega^2 \rightarrow 0}=1\,, \hspace*{1cm}
 \Big(\frac{-\tilde{\bar{P}}^2  }{ (\bar{P} \!\cdot\!\hat{\tilde{q}})^2}
\Big)_{\omega^2 \rightarrow 0}=1\,,
\nonumber \\
&& (\sqrt{D2})_{\omega^2 \rightarrow 0}=2\sqrt{E2}\,
\frac{\bar{P} \ccdot \omega}{|\hat{q} \ccdot \omega| }\,, \hspace*{1cm}
\Big(\frac{2\bar{P} \!\cdot\!\hat{\tilde{q}}}{\sqrt{D0}}
\Big)_{\omega^2 \rightarrow 0}=
-\frac{\hat{q} \ccdot \omega}{|\hat{q} \ccdot \omega|} \,, \hspace*{1cm}
\nonumber \\
&&\Big(\frac{2\bar{P} \!\cdot\!\hat{\tilde{q}}}{\sqrt{D0\,D2}}
\Big)_{\omega^2 \rightarrow 0}=
-\frac{\hat{q} \ccdot \omega}{2\bar{P} \ccdot \omega \sqrt{E2}}\,.
\end{eqnarray} 
The above equalities allow one to make the relation between results presented 
in the previous section and the present one, 
Eqs. (\ref{eq:if3}, \ref{eq:ff3}) for the coefficient $\alpha$, 
Eqs. (\ref{eq:if15}, \ref{eq:ff13}) for the quantities $d$ and $f$,
Eqs. (\ref{eq:if20}, \ref{eq:ff18}) for the quantity $g$, 
Eqs. (\ref{eq:if24}, \ref{eq:ff23}) for the quantity $g'$. 
Relations with the quantities $d$ and $f$ given in Eq. (\ref{eq:lf77}) 
are made by taking the further equality $\hat{q} \ccdot \omega=0$.

Results presented in this section could be used for an approach 
inspired from the Dirac point form based on a hyperboloid surface 
\cite{Desplanques:2004rd}.
Form factors in this approach amount to integrate the above results 
on the orientation of the front, $\vec{\omega}$, with an appropriate weight.

\section{Incorporation of different masses (and spin 1/2) 
in the scalar-constituent results: generalized ``point form"} 
The relation between momenta to start with in the present ``point form" 
\cite{Bakamjian:1961,Sokolov:1978} is given by:
\begin{eqnarray} 
p_{i,f}^{\mu}&\!=\!&
\frac{P_{i,f}^{\mu}}{\sqrt{P^2_{i,f}}}\,\sqrt{s^0_{i,f}}-p^{\mu}\, .
\label{eq:pf1}
\end{eqnarray}
{\bf Derivation of $\alpha$:}\\
One has the following equations:
\begin{eqnarray}
&&(p_i-p_f)^{\mu}=\frac{P_{i}^{\mu}}{\sqrt{P^2_{i}}}\,\sqrt{s^0_{i}}
-\frac{P_{f}^{\mu}}{\sqrt{P^2_{f}}}\,\sqrt{s^0_{f}}\, ,
\nonumber \\
&&(p_i-p_f)^2=\Big( \frac{P_{i}}{\sqrt{P^2_{i}}}\;\sqrt{s^0_i}-
\frac{P_{f}}{\sqrt{P^2_{f}}}\sqrt{s^0_f} \Big)^2\, .
\end{eqnarray}
Incorporating corrections for space-time translation properties, one
successively gets:
\begin{eqnarray}
&&q^2=s_i\!+\!s_f\!-\!2\sqrt{s_i\,s_f}
``\Big(\frac{P_{i}\!\cdot\! P_{f}\cdot}{\sqrt{P^2_{i}\,{P^2_{f}}}}\Big)"\, ,
\nonumber \\ 
&&\frac{s_i\!+\!s_f\!-\!q^2}{2\sqrt{s_i\,s_f}}
=``\Big(\frac{P_i\cdot P_f}{\sqrt{P_i^2\,P_f^2}}\Big)"
 =\frac{M_i^2\!+\!M_f^2\!-\!\alpha^2 q^2}{
 \sqrt{(M_i^2\!+\!M_f^2)^2 \!-\!``(P_i^2\!-\!P_f^2)^2"}} \,,
\nonumber \\
&&\Bigg(\frac{s_i+s_f-q^2}{2\sqrt{s_i\,s_f}}\Bigg)^2
\,\Big(1-\frac{2\alpha^2(\hat{v} \!\cdot\! q)^2}{ M_i^2\!+\!M_f^2}\,
(1-\frac{\alpha^2q^2}{2(M_i^2\!+\!M_f^2)}) \Big)=
(1-\frac{\alpha^2q^2}{M_i^2\!+\!M_f^2})^2\,,
\end{eqnarray}
where $\hat{v}^{\mu}=(P_i+P_f)^{\mu}/\sqrt{(P_i+P_f)^2} $.
The solution of the last equation is given by:
\begin{eqnarray}
\alpha^2&\!=\!&\frac{(M_i^2\!+\!M_f^2)\;D}{  
4s_i\,s_f\!+\!(2\bar{s}\!-\!q^2)^2(\hat{v} \!\cdot\! \hat{q})^2
+(2\bar{s}\!-\!q^2)
\sqrt{4s_i\,s_f\!+\!(2\bar{s}\!-\!q^2)^2(\hat{v} \!\cdot\! \hat{q})^2}
\sqrt{1+(\hat{v} \!\cdot\! \hat{q})^2}}\, .
\nonumber \\
\end{eqnarray}
Guided by a relation obtained elsewhere \cite{Desplanques:2008fg}, we write:
\begin{eqnarray}
\alpha^2&\!=\!&\frac{M_i^2\!+\!M_f^2}{4\sqrt{c_1}\,(\sqrt{c_1}\!+\!\sqrt{c_2})}
\end{eqnarray}
where:
\begin{eqnarray}
&&c_1=\frac{4s_i\,s_f\!+\!(2\bar{s}\!-\!q^2)^2(\hat{v} \!\cdot\! \hat{q})^2}{4D}
=\frac{q^2}{4}+\frac{(2\bar{s}\!-\!q^2)^2}{4D}\,
(1+(\hat{v} \!\cdot\! \hat{q})^2)\,,
\nonumber \\
&&c_2=\frac{(2\bar{s}\!-\!q^2)^2}{4D}(1+(\hat{v} \!\cdot\!
\hat{q})^2)\,,
\nonumber \\
&&c_1\!-\!\frac{(\hat{v} \!\cdot\! q)^2}{4}=\frac{s_i\,s_f}{D}\,
(1+(\hat{v} \!\cdot\! \hat{q})^2)\,.
\end{eqnarray}
Useful relations involving the 4-momenta $P_{i,f}$ are:
\begin{eqnarray}
\frac{``P_i^2+P_f^2"}{2``P_i^2"}=
\frac{1}{1\!-\!\frac{\hat{v} \cdot q}{2\sqrt{c_1}}}\,,\hspace*{1.0cm}
\frac{``P_i^2+P_f^2"}{2``P_f^2"}=
\frac{1}{1\!+\!\frac{\hat{v} \cdot q}{2\sqrt{c_1}}}\,.
\end{eqnarray}
\begin{eqnarray}
``(P_i\!+\!P_f)^2"=2(M_i^2\!+\!M_f^2)-\alpha^2 q^2=
2(M_i^2\!+\!M_f^2)
\Big( 1-\frac{q^2}{8\sqrt{c_1}\,(\sqrt{c_1}+\sqrt{c_2})} \Big)
\nonumber \\
=2(M_i^2\!+\!M_f^2)\Big(1+\frac{\sqrt{c_2}-\sqrt{c_1}}{2\sqrt{c_1}}\Big)
=(M_i^2\!+\!M_f^2)\,\frac{\sqrt{c_2}+\sqrt{c_1}}{\sqrt{c_1}}\, ,\hspace*{1cm}
\nonumber \\
``P^2_f\!-\!P^2_i"=``(P_i\!+\!P_f)\! \cdot\! q " = \hat{v} \!\cdot\! q \;
\sqrt{(M_i^2\!+\!M_f^2)\,\frac{\sqrt{c_2}+\sqrt{c_1}}{\sqrt{c_1}}}
\;\frac{\sqrt{(M_i^2\!+\!M_f^2)}}{
2 \sqrt{\sqrt{c_1}(\sqrt{c_2}+\sqrt{c_1})} }\hspace*{0cm}
\nonumber \\
=\frac{\hat{v} \!\cdot\! q }{2\sqrt{c_1}} \,(M_i^2\!+\!M_f^2)\, .\hspace*{8cm}
\end{eqnarray}

\noindent
{\bf Expressions of $p \!\cdot\! \hat{v}$ and $p \!\cdot\! \hat{q}$ 
in terms of $s_i, \; s_f$}\\
From squaring Eq. (\ref{eq:pf1}), one  gets:
\begin{eqnarray}
``2p \!\cdot\!   P_{i,f}"=
``\sqrt{\frac{P^2_{i,f}}{s^0_{i,f}}} \Big(s^0_{i,f}\!+\!m_2^2\!-\!m_1^2\Big)"\, .
\end{eqnarray}
By separating terms symmetrical and antisymmetrical in the exchange of
initial and final states, one gets:
\begin{eqnarray}
p \!\cdot\! \hat{v}=\frac{1}{2\sqrt{2}\sqrt{\sqrt{c_1}\!+\!\sqrt{c_2}}}
\;\Bigg((1\!+\!\frac{\Delta m^2}{s_f})\sqrt{s_f(\sqrt{c_1}\!+\!\frac{\hat{v} \!\cdot\! q}{2})}
+(1\!+\!\frac{\Delta m^2}{s_i})\sqrt{s_i(\sqrt{c_1}\!-\!\frac{\hat{v} \!\cdot\! q}{2})}
\Bigg)\,,
\nonumber \\
p \!\cdot\! \hat{q}=\frac{1}{2\sqrt{2}\sqrt{\sqrt{c_2}\!-\!\sqrt{c_1}}}
\;\Bigg((1\!+\!\frac{\Delta m^2}{s_f})\sqrt{s_f(\sqrt{c_1}\!+\!\frac{\hat{v} \!\cdot\! q}{2})}
-(1\!+\!\frac{\Delta m^2}{s_i})\sqrt{s_i(\sqrt{c_1}\!-\!\frac{\hat{v} \!\cdot\! q}{2})}
\Bigg)\,.
\label{eq:pf10}
\end{eqnarray}
Useful relations, which involve currents for scalar and spin-1/2 constituents, 
are the following ones:
\begin{eqnarray}
&&``(p_i\!+\!p_f)" \!\cdot\! \hat{v} =
\frac{\sqrt{1\!+\!(\hat{v} \ccdot \hat{q})^2}  }{  
\sqrt{2}\sqrt{\sqrt{c_1}\!+\!\sqrt{c_2}}}
\Bigg (\sqrt{ \frac{s_i}{(\sqrt{c_1}\!-\!\frac{\hat{v} \cdot  q}{2})}}  
+      \sqrt{ \frac{s_f}{(\sqrt{c_1}\!+\!\frac{\hat{v} \cdot  q}{2})}} \Bigg)
\; \frac{2\bar{s}\!-\!2\Delta m^2\!-\!q^2}{2\sqrt{D}}\, ,
\nonumber \\
&&``\Big((p_i\!+\!p_f) \!\cdot\! \hat{v} \,(\bar{s}^0-(m_1\!-\!m_2)^2)
-(p_i\!-\!p_f) \!\cdot\! \hat{v}\,\frac{s^0_i\!-\!s^0_f}{2}
+p \!\cdot\! \hat{v}\, (p_i\!-\!p_f)^2\Big)"
\nonumber \\
&&\hspace*{2cm}=\frac{\sqrt{1\!+\!(\hat{v} \cdot \hat{q})^2}  }{  
\sqrt{2}\sqrt{\sqrt{c_1}\!+\!\sqrt{c_2}}}
\Bigg (\sqrt{ \frac{s_i}{(\sqrt{c_1}\!-\!\frac{\hat{v} \cdot  q}{2})}}  
+      \sqrt{ \frac{s_f}{(\sqrt{c_1}\!+\!\frac{\hat{v} \cdot  q}{2})}} \Bigg)
\nonumber \\
&&\hspace*{3cm} \times \,\frac{2s_is_f-\Delta m^2(2\bar{s}\!-\!q^2) 
-(m_1\!-\!m_2)^2 (2\bar{s}\!-\!2\Delta m^2\!-\!q^2) }{2\sqrt{D}}\, .
\label{eq:pf11}
\end{eqnarray}
It is noticed that, in making the ratio of the two expressions, 
many factors cancel out. The appearance of some of these factors 
is not straigthforward, especially in the second one. 
Details about factorizing some of them may be found in the appendix.

Inverting  Eqs. (\ref{eq:pf10}) to get $s_i, \; s_f$ in terms 
of $p \!\cdot\! \hat{v}$ and $p \!\cdot\! \hat{q}$, 
which could be useful for some particular studies,  seems to be possible 
but the task is considerably more complicated than for the other approaches.
Depending on the way  to deal with the problem, this requires solving 
a third or fourth degree equation and no simple solution was found yet 
(assuming there is one).

\noindent
{\bf Jacobian}\\
As in previous approaches, the calculation of the Jacobian for the transformation
of variables $\vec{p}$ to $s_i,s_f$ and a third variable can be made in two
steps: (i) from $\vec{p}$ to  $p \!\cdot\! \hat{v},\;p \!\cdot\! \hat{q} $ 
and $p^z$ where the $z$ direction is perpendicular to the plane determined by
the 3-vectors $\vec{\hat{v}}$ and  $\vec{\hat{q}}$, and (ii) from these variables 
to $s_i,s_f, p^z$. 

The first step can be made by using Eq. (112) of Ref. \cite{Desplanques:2008fg} 
with  $a^{0,x,y,z}=0,0,0,1$; $b^{\mu}=\hat{v}^{\mu}$; $c^{\mu}=\hat{q}^{\mu}$. 
One then obtains:
\begin{eqnarray}
\frac{d\vec{p}}{e_p}= |J_1| \;
d(p^z)\, d(p\!\cdot\! \hat{v}) \,d(p\!\cdot\! \hat{q})\, ,\label{eq:jac1}
\end{eqnarray}
with: 
\begin{eqnarray}
|J_1|=|(m_2^2 \!+\! p^{z2})(1\!+\! (\hat{v}\ccdot \hat{q})^2)
-2 p \ccdot \hat{q} \;p \ccdot \hat{v} \;\hat{v} \ccdot \hat{q}
-(p \ccdot \hat{v})^2 +(p \ccdot \hat{q})^2|^{-\frac{1}{2}}\,.
\end{eqnarray}
Replacing $p \!\cdot\! \hat{v}, \;p \!\cdot\! \hat{q} $ by
their expression given in Eq. (\ref{eq:pf10}), one can write $|J_1|$ as:
\begin{eqnarray}
|J_1|=\frac{\theta(\cdots)}{\sqrt{1\!+\!(\hat{v} \cdot \hat{q})^2} \; 
\sqrt{\frac{s_is_f}{D}\, c_{\Delta m^2}\!-\!m_2^2\!-\!p^{z2}}}\, .
\end{eqnarray}
The second step is more complicated than in previous approaches as the
quantities $p  \cdot  \hat{v}, \;p  \cdot  \hat{q} $ depend 
on both $\bar{s}$ and $s_i\!-\!s_f$. One obtains:
\begin{eqnarray}
d(p\!\cdot\! \hat{v}) \,d(p\!\cdot\! \hat{q})&=&d\bar{s}\;d(s_i\!-\!s_f) 
\frac{\sqrt{1\!+\!(\hat{v} \ccdot \hat{q})^2}}{4D\sqrt{D}\sqrt{-q^2}}
\nonumber \\   &&\times
(2\bar{s}\!-\!2\Delta m^2\!-\!q^2)
\Bigg (1\!-\!\frac{\Delta m^2 }{2s_is_f} \Big(2\bar{s}+
\frac{\hat{v} \ccdot \hat{q}\;(s_i\!-\!s_f)}{2\sqrt{c_1}}\Big) \Bigg)\,.
\end{eqnarray}
Putting all factors together, we now get:
\begin{eqnarray}
&&\frac{d\vec{p}}{e_p} \; 
=\sum\frac {d\bar{s}\, d(s_i\!-\!s_f)\, dp^z\, \theta(\cdots)
\;(2\bar{s}\!-\!2\Delta m^2\!-\!q^2) 
\Big (1\!-\!\frac{\Delta m^2 }{2s_is_f} \Big(2\bar{s}+
\frac{\hat{v} \cdot \hat{q}\;(s_i\!-\!s_f)}{2\sqrt{c_1}}\Big) \Big)}{  
4QD\sqrt{ D}\,\sqrt{\frac{s_is_f}{D}\, c_{\Delta m^2}\!-\!m_2^2\!-\!p^{z2}} }\, ,
\label{eq:pf15}
\end{eqnarray}
where the sum symbol accounts for the existence of two values of $p_z$ (and of
$e_p$) to be considered.

\noindent
{\bf Charge form factor (scalar constituents)}\\
In the generalized hyperplane approach, the integration volume had 
to be combined with a factor $(``(2p\!+\!p_i\!+\!p_f)"\!\cdot\! \lambda) /
(2``(p_i\!+\!p_f)" \!\cdot\! \lambda) $ to get a Lorentz-invariant result 
(and similarly for the front-form approach). As 
the ``point-form" approach provides naturally Lorentz-invariant results, 
there  is no need to introduce such an extra factor. 
However, another factor is required to recover the appropriate 
normalization, given by Eq. (\ref{eq:dp18}). This role is played 
by the inverse of the last factor at the numerator in Eq. (\ref{eq:pf15}), 
which can be put in the form: 
\begin{eqnarray}
\frac{1}{1\!-\!\frac{\Delta m^2 }{2s_is_f} \Big(2\bar{s}+
\frac{\hat{v} \cdot \hat{q}\;(s_i\!-\!s_f)}{2\sqrt{c_1}}\Big) }
=^{``}\Bigg(\frac{P_i^2\!+\!P_f^2  }{ 
P_i^2 \Big(1\!-\!\frac{\Delta m^2}{s^0_i}\Big )
\!+\!P_f^2\Big(1\!-\!\frac{\Delta m^2}{s^0_f}\Big )}\Bigg)^{"}
\label{eq:pf16}
\end{eqnarray}
In the limit of a zero momentum transfer, and for a system at rest, 
the factor at the r.h.s. reads: 
\begin{eqnarray}
 \frac{ s}{ s-\Delta m^2} =\frac{(e_1+e_2)^2 }{ (e_1+e_2)^2- \Delta m^2} 
 = \frac{e_1+e_2}{ 2e_1} \, ,
\label{eq:pf17}
\end{eqnarray}
which identifies to the needed factor appearing in Eq. (\ref{eq:dp18}). 
It is therefore appropriate to write Eq.  (\ref{eq:pf15}) as follows:
\begin{eqnarray}
&&\hspace*{-1cm}\frac{d\vec{p}}{e_p} \; ^{``}\Bigg (\frac{P_i^2\!+\!P_f^2  }{ 
P_i^2 \Big(1\!-\!\frac{\Delta m^2}{s^0_i}\Big )
\!+\!P_f^2\Big(1\!-\!\frac{\Delta m^2}{s^0_f}\Big )}\Bigg)^{"}
=\sum\frac {d\bar{s}\, d(s_i\!-\!s_f)\, dp^z\, \theta(\cdots)
\;(2\bar{s}\!-\!2\Delta m^2\!-\!q^2) }{  
4QD\sqrt{ D}\,\sqrt{\frac{s_is_f}{D}\, c_{\Delta m^2}\!-\!m_2^2\!-\!p^{z2}} }\, ,
\label{eq:pf18}
\end{eqnarray}
By inserting this equality in the form-factor expression, one gets:
\begin{eqnarray} 
``F_1(Q^2)"&\!=\!&\frac{16\pi^2}{N}\! \int\! \frac{d\vec{p}}{(2\pi)^3}\;
\frac{1}{e_p}    \;
^{``}\Bigg(\frac{P_i^2\!+\!P_f^2  }{ 
P_i^2 \Big(1\!-\!\frac{\Delta m^2}{s^0_i}\Big )
\!+\!P_f^2\Big(1\!-\!\frac{\Delta m^2}{s^0_f}\Big )}\;\tilde{\phi}(\vec{k_f}^2)\;\tilde{\phi}(\vec{k_i}^2) \Bigg)^{"}
\nonumber \\
&\!=\!&\frac{2}{\pi N}\!\int \! \! \int d\bar{s} \; d(\frac{s_i\!-\!s_f}{Q}) \,
\phi(s_f)\, \phi(s_i) \; \frac{ \theta(\cdots) }{4\sqrt{D}} \,
\Big [\frac{2\bar{s}\!-\! 2\Delta m^2\!+\!Q^2 }{ D}   \Big] 
 \nonumber \\
&& \hspace*{0.6cm} \times
\sum \int \!\frac {dp^z  }{ 
\sqrt {\frac {s_i\,s_f}{D }\,c_{\Delta m^2}-m_2^2-p^{z2}}   }  
\nonumber \\
&\!=\!&\frac{1}{N}\!\int \! \! \int d\bar{s} \; d(\frac{s_i\!-\!s_f}{Q}) \;
\phi(s_f)\;\phi(s_i) \;
\frac{ [2\bar{s}\!-\! 2\Delta m^2\!+\!Q^2 ]\,\theta(\cdots)  }{D\sqrt{D}} \, .
\label{eq:pf19}
\end{eqnarray} 
One then recovers the expression for the form factor given 
by Eq. (\ref{eq:dp12}).

\noindent
{\bf Charge form factors (spin-1/2 constituents)}\\
To account for the spin-1/2 nature of the constituents, one can introduce 
in  the integrand for the scalar-constituent case the ratio 
of the corresponding matrix elements of the current, 
$I_v=I \cdot \hat{v}$ 
and $\tilde{I}_v =\tilde{I}\cdot \hat{v} $:
\begin{eqnarray}
&& \frac{I_v}{\tilde{I}_v}=``\Bigg(
 \frac{(p_i\!+\!p_f)\!\cdot\! \hat{v}\;\Big(\bar{s}^0\!-\!(m_1\!-\!m_2)^2\Big)
\!-\!(p_i\!-\!p_f)\!\cdot\! \hat{v}\,\frac{s^0_i\!-\!s^0_f}{2} 
+p\!\cdot\!  \hat{v}\,(p_i\!-\!p_f)^2
 }{(p_i\!+\!p_f)\!\cdot\! \hat{v}
\;\sqrt{s^0_i\!-\!(m_1\!-\!m_2)^2}\;\sqrt{s^0_f\!-\!(m_1\!-\!m_2)^2}}\Bigg)"
 \nonumber \\
&& \hspace*{0.6cm}=
\frac{2s_is_f-\Delta m^2(2\bar{s}\!-\!q^2) 
-(m1\!-\!m2)^2 (2\bar{s}\!-\!2\Delta m^2\!-\!q^2)   }{  
(2\bar{s}\!-\!2\Delta m^2\!-\!q^2)
\sqrt{s_i\!-\!(m_1\!-\!m_2)^2}\;\sqrt{s_f\!-\!(m_1\!-\!m_2)^2}\;}\,.
\label{eq:pf22}
\end{eqnarray}
Introducing the above ratios in Eq. (\ref{eq:pf19}), one gets the expression
of the charge form factor for  spin-1/2 constituents:
\begin{eqnarray} 
``F_1(Q^2)"\!&\!=\!&\!\frac{2}{\pi N}\!\int \! \! \int \! d\bar{s} \, 
d(\frac{s_i\!-\!s_f}{Q}) \, \phi(s_f) \, \phi(s_i) \,
   \frac{ \theta(\cdots)}{4\sqrt{D}}\,
  \frac{  \Big [\frac{2s_is_f\!-\!\Delta m^2(2\bar{s}\!+\!Q^2) 
\!-\!(m_1\!-\!m_2)^2 (2\bar{s}\!-\!2\Delta m^2\!+\!Q^2)}{ D}  \Big] }{ 
  \sqrt{s_i\!-\!(m_1\!-\!m_2)^2}\;\sqrt{s_f\!-\!(m_1\!-\!m_2)^2}} 
\nonumber \\
&& \hspace*{0.6cm} \times 
\sum \int \!\frac { dp^z  }{ 
\sqrt {\frac {s_i\,s_f}{D }\,c_{\Delta m^2}-m_2^2-p^{z2} } }  
\nonumber \\
&\!=\!&\!\frac{1}{N}\!\int \! \! \int d\bar{s} \; d(\frac{s_i\!-\!s_f}{Q}) \;
\phi(s_f)\;\phi(s_i)
\nonumber \\
&& \times
 \frac{ \Big [2s_is_f\!-\!\Delta m^2(2\bar{s}\!+\!Q^2) 
\!-\!(m_1\!-\!m_2)^2 (2\bar{s}\!-\!2\Delta m^2\!+\!Q^2)\Big]\;\theta(\cdots)}{
D\sqrt{D}\;\sqrt{s_i\!-\!(m_1\!-\!m_2)^2}\;\sqrt{s_f\!-\!(m_1\!-\!m_2)^2} } \, .
\label{eq:pf23}
\end{eqnarray} 
%

\section{Conclusion}
We have extended a previous work dealing with constraints from space-time
translations for the calculation of form factors of $J=0$ systems 
composed of scalar constituents. We both considered the case of unequal-mass and
spin-1/2  constituents, which are physically more relevant. 
As in the previous work, we were able to show that accounting for the above
constraints could lead to the same results for form factors calculated in
different forms, using the same solution of a mass operator. 
We stress that this important result supposes the introduction 
of two-body currents to all orders in the interaction, 
which, actually, was done by a modification of the ingredients 
entering the one-body contribution (wave function and current). 
The two-body currents differ from those considered elsewhere 
\cite{Desplanques:2003nk,Desplanques:2009}, which were aimed 
to obtain the right asymptotic behavior of form factors. 
Results for form factors were also shown to be the same as those obtained 
in a dispersion-relation approach. The equivalence supposes 
an appropriate current. At $Q^2=0$, where constraints for transformations of
currents under space-time translations have no effect, the current has to be
chosen in such a way that  invariance of form factors under rotation and boost 
is fulfilled. At finite values of $Q^2$, at least for the charge form factor, 
the expression of the current is not unexpected, probably because it has 
some relationship with a conserved one. Only the ``point form" 
requires  the introduction of a factor that could not be anticipated. 
All results satisfy properties pertinent to the Poincar\'e group: 
invariance under rotations, invariance under boosts 
and constraints from space-time translations. 
These somewhat geometrical properties do not imply however that physics 
is fully accounted for. For the  pion and kaon form factors \cite{He:2004ba}, 
which present results could be applied to, extra specific two-body currents, 
mentioned above, are definitively needed \cite{Desplanques:2009}.

Present results should be extended to other systems with more than 
two constituents or a non-zero spin. The first one could be easily done by 
identifying the squared mass $m_2^2$ with the invariant squared mass
of the spectator particles, $(p_2+p_3+\cdots)^2$ .
The consideration of non-zero spin systems could be more complicated. 
The current matrix element then implies dependence on powers of $q^{\mu}$ 
beside the scalar product $q^2$. This supposes that some of the components 
should be treated on a different footing, instead of a uniform one 
as done in this work. As there is some freedom in implementing 
the constraints from space-time translations for form factors depending 
exclusively on $q^2$, we believe that this freedom could be used in the case 
where the current matrix element also depends on $q^{\mu}$. The implementation of the 
constraints may not be so easy for such terms as for $q^2$.
Another but different possible extension concerns the time-like region. With this respect, 
we notice that results from the standard front-form approach  
with $q^+=0$, which are unaffected by constraints from space-time translations 
considered in this work, would loose their particular status. 
Indeed, in the time-like region, it is not possible to fulfill 
the condition $q^+=0$. This could lead to extra interesting developments.

{\bf Acknowledgements}\\
We are very grateful to Dr S. Simula for calling our attention to further
extensions of this work like the case of time-like momentum transfers. 
This work is supported by the National Sciences Foundations No. 10775148 
and by the grant No. KJCX3-SYW-N2.

\appendix

\section{``Point-form'' approach: useful relations ($\Delta m^2=0$ 
and $\Delta m^2 \neq 0$)}
$\bullet$ {\bf Further relations for the case with $\Delta m^2=0$ }\\
We first complete results given in Ref. \cite{Desplanques:2008fg} 
for the case $\Delta m^2=0$  but with $\hat{v} \cdot q \neq 0$. 
In this case, it is still possible to express the quantities $s_i$ and $s_f$ 
in terms of $ p \!\cdot\! \hat{v}$ and $p \!\cdot\! \hat{q}$ with the 
result:
\begin{eqnarray}
s_i=\frac{2}{\sqrt{c_1}\!-\!\frac{\hat{v} \cdot q}{2}}\;
\Big( p \!\cdot\! \hat{v} \sqrt{\sqrt{c_1}\!+\!\sqrt{c_2}}
-p \!\cdot\! \hat{q} \sqrt{\sqrt{c_2}\!-\!\sqrt{c_1}} \Big)^2\,,
\nonumber \\
s_f=\frac{2}{\sqrt{c_1}\!+\!\frac{\hat{v} \cdot q}{2}}\;
\Big( p \!\cdot\! \hat{v} \sqrt{\sqrt{c_1}\!+\!\sqrt{c_2}}
+p \!\cdot\! \hat{q} \sqrt{\sqrt{c_2}\!-\!\sqrt{c_1}} \Big)^2\,.
\end{eqnarray}
These results allow one to get relations that show 
the equivalence of two writings for $\alpha^2$, in terms 
of the variables $s_i,\;s_f$ on the one hand, 
and $p \!\cdot\! \hat{v}$ and $p \!\cdot\! \hat{q}$ on the other hand:
\begin{eqnarray}
&&c_1=\frac{4s_i\,s_f\!+\!(2\bar{s}\!-\!q^2)^2(\hat{v} \!\cdot\! \hat{q})^2}{4D}
=(p \!\cdot\! \hat{v})^2 \!-\! (p \!\cdot\! \hat{q})^2
\!+\!2p \!\cdot\! \hat{v} \,p \!\cdot\! \hat{q} \,\hat{v} \!\cdot\! \hat{q}
\!+\!\frac{(\hat{v} \!\cdot\! q)^2}{4}\, ,
\nonumber \\
&&c_2=\frac{(2\bar{s}\!-\!q^2)^2}{4D}(1+(\hat{v} \!\cdot\!
\hat{q})^2)=(p \!\cdot\! \hat{v})^2 \!-\! (p \!\cdot\! \hat{q})^2
\!+\!2p \!\cdot\! \hat{v} \,p \!\cdot\! \hat{q} \,\hat{v} \!\cdot\! \hat{q}
\!+\!\frac{(\hat{v} \!\cdot\! q)^2}{4}\!-\!\frac{q^2}{4}\,,
\nonumber \\
&&c_1\!-\!\frac{(\hat{v} \!\cdot\! q)^2}{4}=\frac{s_i\,s_f}{D}\,
(1+(\hat{v} \!\cdot\! \hat{q})^2)=
(p \!\cdot\! \hat{v})^2 \!-\! (p \!\cdot\! \hat{q})^2
\!+\!2p \!\cdot\! \hat{v} \,p \!\cdot\! \hat{q} \,\hat{v} \!\cdot\! \hat{q}
\, . 
\end{eqnarray}

\noindent
$\bullet$ {\bf Relations for the case $\Delta m^2 \neq 0$ }\\
We give here details that are relevant to the factorization of a common factor 
in quantities given in the main text, Eq. (\ref{eq:pf11}). 
For this case, we need  expressions of $(p_i\!+\!p_f)\!\cdot\!\hat{v}$ 
and  $(p_i\!-\!p_f)\!\cdot\!\hat{v}$, to be determined,  
and the expression of $p \!\cdot\! \hat{v}$, given in Eq. (\ref{eq:pf10}). 
The above 3 terms have to be multiplied respectively 
by $\frac{s_i\!+\!s_f}{2}\!-\!(m_1\!-\!m_2)^2$,  
$-\frac{s_i\!-\!s_f}{2}$, and $q^2$ for the spin-1/2 constituent case 
and the first one by $\sqrt{(s_i\!-\!(m_1\!-\!m_2)^2)\,(s_f\!-\!(m_1\!-\!m_2)^2)}$ 
for the spin-0 one.

The expression of $(p_i\!+\!p_f) \cdot \hat{v}$  results 
from summing contributions proportional to $(P_i\!+\!P_f) \cdot \hat{v}$, 
$(P_i\!-\!P_f) \cdot \hat{v}$ and $p \cdot \hat{v}$:
\begin{eqnarray}
&& \hspace*{-1cm}``(p_i\!+\!p_f)\!\cdot\!\hat{v}"=\frac{1}{\sqrt{2}}\Bigg( 
\frac{\sqrt{s_i}}{\sqrt{\sqrt{c_1}\!-\!\frac{\hat{v} \cdot q}{2}}} 
+ \frac{\sqrt{s_f}}{\sqrt{\sqrt{c_1}\!+\!\frac{\hat{v} \cdot q}{2}}}\Bigg)
\sqrt{\sqrt{c_1}+\sqrt{c_2}} 
\nonumber \\
&& \hspace*{1cm}-\frac{\hat{v}\!\cdot\!q}{2\sqrt{2}}\Bigg( 
\frac{\sqrt{s_i}}{\sqrt{\sqrt{c_1}\!-\!\frac{\hat{v} \cdot q}{2}}} 
- \frac{\sqrt{s_f}}{\sqrt{\sqrt{c_1}\!+\!\frac{\hat{v} \cdot q}{2}}}\Bigg)
\frac{1}{\sqrt{\sqrt{c_1}+\sqrt{c_2}}}
\nonumber \\
&& \hspace*{1cm}-\frac{1}{\sqrt{2}}\Bigg(
(1\!+\!\frac{\Delta m^2}{s_i})\sqrt{s_i\,(\sqrt{c_1}\!-\!\frac{\hat{v} \!\cdot\! q}{2})}
+(1\!+\!\frac{\Delta m^2}{s_f})\sqrt{s_f\,(\sqrt{c_1}\!+\!\frac{\hat{v} \!\cdot\! q}{2})}
\Bigg) \frac{1}{\sqrt{\sqrt{c_1}+\sqrt{c_2}}}
\nonumber \\
&& \hspace*{0cm}=\frac{1}{\sqrt{2}}\Bigg( 
\frac{\sqrt{s_i}}{\sqrt{\sqrt{c_1}\!-\!\frac{\hat{v} \cdot q}{2}}} 
+ \frac{\sqrt{s_f}}{\sqrt{\sqrt{c_1}\!+\!\frac{\hat{v} \cdot q}{2}}}\Bigg)
\frac{\sqrt{c_1}+\sqrt{c_2}-\sqrt{c_1}
-\Delta m^2 \frac{\sqrt{4c_1\!-\!(\hat{v} \!\cdot\! q)^2} }{ 
\sqrt{4s_is_f}}}{\sqrt{\sqrt{c_1}+\sqrt{c_2}}}
\nonumber \\
&& \hspace*{0cm}=\frac{1}{\sqrt{2}}\Bigg( 
\frac{\sqrt{s_i}}{\sqrt{\sqrt{c_1}\!-\!\frac{\hat{v} \cdot q}{2}}} 
+ \frac{\sqrt{s_f}}{\sqrt{\sqrt{c_1}\!+\!\frac{\hat{v} \cdot q}{2}}}\Bigg)
\frac{\sqrt{c_2}\!-\!\Delta m^2 \frac{\sqrt{4c_1\!-\!(\hat{v} \!\cdot\! q)^2} }{ 
\sqrt{4s_is_f}}}{\sqrt{\sqrt{c_1}+\sqrt{c_2}}}\, 
\nonumber \\
&& \hspace*{0cm}=\frac{1}{\sqrt{2}}\Bigg( 
\frac{\sqrt{s_i}}{\sqrt{\sqrt{c_1}\!-\!\frac{\hat{v} \cdot q}{2}}} 
+ \frac{\sqrt{s_f}}{\sqrt{\sqrt{c_1}\!+\!\frac{\hat{v} \cdot q}{2}}}\Bigg)
\frac{(2\bar{s}\!-\! 2 \Delta m^2\!-\!q^2) 
\sqrt{1+(\hat{v} \!\cdot\! \hat{q})^2}}{2\sqrt{D}\; \sqrt{\sqrt{c_1}+\sqrt{c_2}}}
\, .
\end{eqnarray}

The expression of $(p_i\!-\!p_f) \cdot \hat{v}$  results 
from summing contributions proportional to $(P_i\!+\!P_f) \cdot \hat{v}$  
and $(P_i\!-\!P_f) \cdot \hat{v}$:
\begin{eqnarray}
&&``(p_i\!-\!p_f)\!\cdot\!\hat{v}"=\frac{1}{\sqrt{2}}\Bigg( 
\frac{\sqrt{s_i}}{\sqrt{\sqrt{c_1}\!-\!\frac{\hat{v} \cdot q}{2}}} 
- \frac{\sqrt{s_f}}{\sqrt{\sqrt{c_1}\!+\!\frac{\hat{v} \cdot q}{2}}}\Bigg)
\sqrt{\sqrt{c_1}+\sqrt{c_2}}
\nonumber \\
&& \hspace*{3cm}-\frac{\hat{v}\!\cdot\!q}{2\sqrt{2}}\Bigg( 
\frac{\sqrt{s_i}}{\sqrt{\sqrt{c_1}\!-\!\frac{\hat{v} \cdot q}{2}}} 
+ \frac{\sqrt{s_f}}{\sqrt{\sqrt{c_1}\!+\!\frac{\hat{v} \cdot q}{2}}}\Bigg)
\frac{1}{\sqrt{\sqrt{c_1}+\sqrt{c_2}}}\, .
\end{eqnarray}

Taking into account the expressions for the individual terms given above or in
the main text, the sum of the contributions for the spin-1/2 constituent case 
reads:
\begin{eqnarray}
&&\hspace*{-0.7cm}``(p_i\!+\!p_f)\!\cdot\!\hat{v}"\,\Big(\frac{s_i\!+\!s_f}{2}\!-\!(m_1\!-\!m_2)^2\Big)
-``(p_i\!-\!p_f)\!\cdot\!\hat{v}"\,\frac{s_i\!-\!s_f}{2}
+``p\!\cdot\!\hat{v}"\,q^2 \hspace*{0.9cm}
\nonumber \\
&&
\hspace*{0.1cm}=\frac{1}{\sqrt{2}\sqrt{\sqrt{c_1}\!+\!\sqrt{c_2}}}\;
 \Bigg (\frac{\sqrt{s_i}}{\sqrt{\sqrt{c_1}\!-\!\frac{\hat{v} \cdot q}{2}}} 
+ \frac{\sqrt{s_f}}{\sqrt{\sqrt{c_1}\!+\!\frac{\hat{v} \cdot q}{2}}}\Bigg)
\nonumber \\
&& \hspace*{1.1cm} \times 
\Bigg ( \Big (\sqrt{c_2}\!-\!\Delta m^2 
\frac{\sqrt{4c_1\!-\!(\hat{v} \!\cdot\! q)^2} }{ \sqrt{4s_is_f}}\Big )\,
\Big(\frac{s_i\!+\!s_f}{2}\!-\!(m_1\!-\!m_2)^2\Big) +\frac{\hat{v} \!\cdot\! q}{2}
\,\frac{s_i\!-\!s_f}{2}
\nonumber \\
&& \hspace*{2.1cm}+\Big(\sqrt{c_1}+ 
\Delta m^2 \frac{\sqrt{4c_1\!-\!(\hat{v} \!\cdot\! q)^2} }{ 
\sqrt{4s_is_f}} \Big)\,\frac{q^2}{2}\Bigg)
\nonumber \\
&&\hspace*{0.6cm}-\frac{1}{\sqrt{2}\sqrt{\sqrt{c_1}\!+\!\sqrt{c_2}}}\;\Bigg ( 
\frac{\sqrt{s_i}}{\sqrt{\sqrt{c_1}\!-\!\frac{\hat{v} \cdot q}{2}}} 
- \frac{\sqrt{s_f}}{\sqrt{\sqrt{c_1}\!+\!\frac{\hat{v} \cdot q}{2}}}\Bigg)
\Big((\sqrt{c_1}\!+\!\sqrt{c_2})\frac{s_i\!-\!s_f}{2}
+\frac{\hat{v} \!\cdot\! q}{4}\,q^2\Big)
\nonumber \\
&&\hspace*{0.1cm}=\frac{1}{\sqrt{2}\sqrt{\sqrt{c_1}\!+\!\sqrt{c_2}}}
\Bigg (\frac{\sqrt{s_i}}{\sqrt{\sqrt{c_1}\!-\!\frac{\hat{v} \cdot q}{2}}} 
+ \frac{\sqrt{s_f}}{\sqrt{\sqrt{c_1}\!+\!\frac{\hat{v} \cdot q}{2}}}\Bigg)
\nonumber \\ 
&& \hspace*{1.1cm} \times
\Bigg [ \sqrt{c_2}\,\frac{s_i\!+\!s_f}{2} +\frac{\hat{v} \!\cdot\! q}{2}
\,\frac{s_i\!-\!s_f}{2}+\sqrt{c_1}\,\frac{q^2}{2}  
-\Bigg (\sqrt{s_i\,(1\!+\!\frac{\hat{v} \!\cdot\! q}{2\sqrt{c_1}})} 
  -  \sqrt{s_f\,(1\!-\!\frac{\hat{v} \!\cdot\! q}{2\sqrt{c_1}})}  \Bigg)
  \nonumber \\ 
  &&\hspace*{1.1cm}\times
\Bigg (\frac{\sqrt{c_1}}{2}
\Bigg ( \sqrt{s_i(1\!+\!\frac{\hat{v} \!\cdot\! q}{2\sqrt{c_1}})} 
     -  \sqrt{s_f(1\!-\!\frac{\hat{v} \!\cdot\! q}{2\sqrt{c_1}})}\Bigg)
  + \frac{\sqrt{c_2}}{2}
  \Bigg ( \sqrt{\frac{s_i}{1\!+\!\frac{\hat{v} \cdot q}{2\sqrt{c_1}}}} 
  -\sqrt{\frac{s_f}{1\!-\!\frac{\hat{v} \cdot q}{2\sqrt{c_1}}}}\Bigg)\Bigg)
\nonumber \\ 
  &&\hspace*{1.1cm} -\Delta m^2 
\frac{\sqrt{4c_1\!-\!(\hat{v} \!\cdot\! q)^2} }{ \sqrt{4s_is_f}}\Big )\,
\Big(\frac{s_i\!+\!s_f}{2} -\frac{q^2}{2} \Big) 
\!-\!(m_1\!-\!m_2)^2 \Big (\sqrt{c_2}\!-\!\Delta m^2
\frac{\sqrt{4c_1\!-\!(\hat{v} \!\cdot\! q)^2} }{ \sqrt{4s_is_f}}\Big)
\Bigg]
\nonumber \\
&&\hspace*{0.1cm}=\frac{1}{\sqrt{2}\sqrt{\sqrt{c_1}\!+\!\sqrt{c_2}}}
\Bigg (
\frac{\sqrt{s_i}}{\sqrt{\sqrt{c_1}\!-\!\frac{\hat{v} \cdot q}{2}}} 
+ \frac{\sqrt{s_f}}{\sqrt{\sqrt{c_1}\!+\!\frac{\hat{v} \cdot q}{2}}}\Bigg)
\nonumber \\ 
&& \hspace*{1.1cm} \times
\Bigg [ \sqrt{c_2}\,\frac{s_i\!+\!s_f}{2} +\frac{\hat{v} \!\cdot\! q}{2}
\,\frac{s_i\!-\!s_f}{2}+\sqrt{c_1}\,\frac{q^2}{2}
\nonumber \\
&&\hspace*{1.1cm}- \sqrt{c_1}\,\frac{s_i\!+\!s_f}{2} 
-\frac{\hat{v} \!\cdot\! q}{2}\,\frac{s_i\!-\!s_f}{2}
+\sqrt{s_i\,s_f}\,\sqrt{1-\frac{(\hat{v} \!\cdot\! q)^2}{q^2}}
\frac{\sqrt{s_i\,s_f}}{\sqrt{D}}
\nonumber \\
&& \hspace*{1.1cm} -\sqrt{c_2}\,\frac{s_i\!+\!s_f}{2}+ \sqrt{c1}\,\sqrt{s_i\,s_f}
\frac{\sqrt{D}}{2\sqrt{s_i\,s_f}}
\frac{(2\bar{s}-q^2)}{\sqrt{D}}
\nonumber \\ 
  &&\hspace*{1.1cm} -\Delta m^2 
\frac{\sqrt{4c_1\!-\!(\hat{v} \!\cdot\! q)^2} }{ \sqrt{4s_is_f}}\Big )\,
\Big(\frac{s_i\!+\!s_f}{2} -\frac{q^2}{2} \Big) 
\!-\!(m_1\!-\!m_2)^2 \Big (\sqrt{c_2}\!-\!\Delta m^2
\frac{\sqrt{4c_1\!-\!(\hat{v} \!\cdot\! q)^2} }{ \sqrt{4s_is_f}}\Big)
\Bigg]
\nonumber \\
&&\hspace*{0.1cm}=\frac{\sqrt{1\!+\!(\hat{v} \!\cdot\! \hat{q})^2}}{\sqrt{2}\sqrt{\sqrt{c_1}\!+\!\sqrt{c_2}}}
\Bigg (
\frac{\sqrt{s_i}}{\sqrt{\sqrt{c_1}\!-\!\frac{\hat{v} \cdot q}{2}}} 
+ \frac{\sqrt{s_f}}{\sqrt{\sqrt{c_1}\!+\!\frac{\hat{v} \cdot q}{2}}}\Bigg)
\nonumber \\
&&\hspace*{1.1cm} \times\;\frac{2s_i\,s_f\!-\!\Delta m^2(2\bar{s}\!-\!q^2) 
\!-\!(m_1\!-\!m_2)^2 (2\bar{s}\!-\!2\Delta m^2\!-\!q^2)}{2\sqrt{D}} \,.
\end{eqnarray}
To make the front factor to appear everywhere, we used the relation:
\begin{eqnarray}
(\sqrt{c_1}\!+\!\sqrt{c_2}) \frac{s_i\!-\!s_f}{2}+
\frac{\hat{v} \!\cdot\! q \;q^2}{4}=
\Bigg (\sqrt{s_i\,(1\!+\!\frac{\hat{v} \!\cdot\! q}{2\sqrt{c_1}})} 
  +  \sqrt{s_f\,(1\!-\!\frac{\hat{v} \!\cdot\! q}{2\sqrt{c_1}})}  \Bigg)
  \hspace*{2.1cm}
\nonumber \\ \times
\Bigg [\frac{\sqrt{c_1}}{2}
\Bigg( \sqrt{s_i(1\!+\!\frac{\hat{v} \!\cdot\! q}{2\sqrt{c_1}})} 
     -  \sqrt{s_f(1\!-\!\frac{\hat{v} \!\cdot\! q}{2\sqrt{c_1}})}\Bigg)
  + \frac{\sqrt{c_2}}{2}
  \Bigg( \sqrt{\frac{s_i}{1\!+\!\frac{\hat{v} \cdot q}{2\sqrt{c_1}}}} 
  -\sqrt{\frac{s_f}{1\!-\!\frac{\hat{v} \cdot q}{2\sqrt{c_1}}}}\Bigg)
   \Bigg]\,.
\end{eqnarray}
The corresponding result for the spin-0-constituent case is:
\begin{eqnarray}
&&\sqrt{(s_i\!-\!(m_1\!-\!m_2)^2)\,(s_f\!-\!(m_1\!-\!m_2)^2)}\;\; 
``(p_i\!+\!p_f)\!\cdot\!\hat{v}"
\nonumber \\
&&\hspace*{2cm}
=\frac{1}{\sqrt{2}\sqrt{\sqrt{c_1}\!+\!\sqrt{c_2}}}\Bigg (
\frac{\sqrt{s_i}}{\sqrt{\sqrt{c_1}\!-\!\frac{\hat{v} \cdot q}{2}}} 
+ \frac{\sqrt{s_f}}{\sqrt{\sqrt{c_1}\!+\!\frac{\hat{v} \cdot q}{2}}}\Bigg)
\nonumber \\
&&\hspace*{3cm} \times \sqrt{(s_i\!-\!(m_1\!-\!m_2)^2)\,(s_f\!-\!(m_1\!-\!m_2)^2)}\;
\Big (\sqrt{c_2}\!-\!\Delta m^2 \frac{\sqrt{4c_1\!-\!(\hat{v} \!\cdot\! q)^2} }{ 
\sqrt{4s_is_f}} \Big)
\nonumber \\
&&\hspace*{2cm}=\frac{\sqrt{1\!+\!(\hat{v} \!\cdot\! \hat{q})^2}}{\sqrt{2}\sqrt{\sqrt{c_1}\!+\!\sqrt{c_2}}}
\Bigg (
\frac{\sqrt{s_i}}{\sqrt{\sqrt{c_1}\!-\!\frac{\hat{v} \cdot q}{2}}} 
+ \frac{\sqrt{s_f}}{\sqrt{\sqrt{c_1}\!+\!\frac{\hat{v} \cdot q}{2}}}\Bigg)\;
\nonumber \\
&&\hspace*{3cm}\times \frac{\sqrt{(s_i\!-\!(m_1\!-\!m_2)^2)\,(s_f\!-\!(m_1\!-\!m_2)^2)}
\;(2\bar{s}\!-\! 2 \Delta m^2\!-\! q^2)}{2\sqrt{D}}\,.
\end{eqnarray}
%


\end{document}